\documentclass[10pt,letterpaper]{article}
\usepackage[top=0.85in,left=2.75in,footskip=0.75in]{geometry}

\usepackage{amsmath,amssymb}
\usepackage{changepage}
\usepackage[utf8x]{inputenc}
\usepackage{textcomp,marvosym}
\usepackage{cite}
\usepackage{nameref,hyperref}
\usepackage[right]{lineno}
\usepackage{microtype}
\DisableLigatures[f]{encoding = *, family = * }
\usepackage[table]{xcolor}
\usepackage{array}
\newcolumntype{+}{!{\vrule width 2pt}}
\newlength\savedwidth

\newcommand\thickhline{\noalign{\global\savedwidth\arrayrulewidth\global\arrayrulewidth 2pt}%
\hline
\noalign{\global\arrayrulewidth\savedwidth}}
\raggedright
\usepackage[aboveskip=1pt,labelfont=bf,labelsep=period,justification=raggedright,singlelinecheck=off]{caption}

\makeatletter
\renewcommand{\@biblabel}[1]{\quad#1.}
\makeatother
\usepackage{lastpage,fancyhdr,graphicx}
\usepackage{epstopdf}
\fancyheadoffset[L]{2.25in}
\fancyfootoffset[L]{2.25in}
\usepackage{multirow}
\usepackage{todonotes}
\usepackage{multicol}
\usepackage{url}

\usepackage{array}
\newcommand{\PreserveBackslash}[1]{\let\temp=\\#1\let\\=\temp}
\newcolumntype{C}[1]{>{\PreserveBackslash\centering}p{#1}}
\newcolumntype{L}[1]{>{\PreserveBackslash\raggedright}p{#1}}

\usepackage{algpseudocode}
\usepackage{algorithm}
\usepackage[normalem]{ulem}

\begin{document}
\vspace*{0.2in}

\begin{flushleft}
{\Large
\textbf\newline{Improved prediction of hiking speeds using a data driven approach}
}
\newline
\\
Andrew Wood\textsuperscript{1*},
William Mackaness\textsuperscript{2},
T. Ian Simpson\textsuperscript{1},
J. Douglas Armstrong\textsuperscript{1}
\\

\bigskip
\textbf{1} School of Informatics, University of Edinburgh, Edinburgh, UK
\\
\textbf{2} School of Geoscience, University of Edinburgh, Edinburgh, UK
\\
\bigskip
* andrew.wood@ed.ac.uk
\end{flushleft}

\section*{Abstract}
Hikers and hillwalkers typically use the gradient in the direction of travel (walking slope) as the main variable in established methods for predicting walking time (via the walking speed) along a route. Research into fell-running has suggested further variables which may improve speed algorithms in this context; the gradient of the terrain (hill slope) and the level of terrain obstruction. Recent improvements in data availability, as well as widespread use of GPS tracking now make it possible to explore these variables in a walking speed model at a sufficient scale to test statistical significance. We tested various established models used to predict walking speed against public GPS data from almost 88,000 km of UK walking / hiking tracks. Tracks were filtered to remove breaks and non-walking sections. A new generalised linear model (GLM) was then used to predict walking speeds. Key differences between the GLM and established rules were that the GLM considered the gradient of the terrain (hill slope) irrespective of walking slope, as well as the terrain type and level of terrain obstruction in off-road travel. All of these factors were shown to be highly significant, and this is supported by a lower root-mean-square-error compared to existing functions. We also observed an increase in RMSE between the GLM and established methods as hill slope increases, further supporting the importance of this variable. 
%\section*{Author summary}
% Please keep the Author Summary between 150 and 200 words
% Use first person. PLOS ONE authors please skip this step. 
% Author Summary not valid for PLOS ONE submissions.  

\nolinenumbers
\section*{Introduction}
\label{0Background}
Knowing how fast people are able to walk between locations is critical information in many situations. In hiking and hillwalking scenarios, this information is vital for safety reasons. If you are leaving in the morning for a hike then it is good practice to provide an estimated return time such that emergency services can be contacted if you get into difficulty and do not return \cite{Mountaineering2023Route}. An inaccurate estimate for how long a route will take could lead to unnecessary callouts, or delay a callout in a situation where every minute is important. Furthermore, in circumstances where a hiker has gone missing, an accurate measure of walking speed can help to restrict a potential search area around a last known location. Finally, when out on a hike there are situations where hikers may be deciding whether to follow a footpath, or take a more direct cross-country route. Accurate estimates of the walking speed and time for both scenarios are required to be able to select the optimal route.

There are a multitude of factors which can impact the walking speed and time predictions for a route \cite{Mountaineering2023Planning}, although these can generally be split into two categories \cite{Miao2023AnalysisEnvironment, Liang2020HowAreas}. The first category covers the individual effects which depend on who precisely is undertaking the walk, and when they are doing it. These effects include group size (larger groups often walk slower), age or fitness of  participants, and weather conditions, as well as the aim of the walk (afternoon stroll vs. specific hike). The second category covers the fixed effects which will affect all individuals who attempt the same route. These include how steep the terrain is and whether the route is paved, along a track or in wild country.

Most of the individual effects cannot be modelled without considerable prior knowledge about the person who is planning a route. Therefore, most existing hiking route planners calculate the walking speed solely based on the terrain, and this is presented as the average time (or time range) it takes to complete a hike. It is then left up to the individual to tune the predicted time for a hike given their knowledge about personal ability and circumstances.

Formulae of varying complexity have been proposed to estimate human walking speed and time along a projected path. A popular early method that is still widely used was put forward by Naismith \cite{Naismith1892CruachMore} which calculates walking time under normal conditions as:
\begin{quote}
``\textit{an hour for every three miles on the map, with an additional hour for every 2,000 feet of ascent.}''
\end{quote}
This approximates to a walking speed of 5 km/h with 10 minutes added on for every 100 m of ascent. This was later adjusted by Aitken \cite{Aitken1977WildernessScotland}, who introduced a reduced base movement speed of 4 km/h on surfaces which are not paths or roads. Naismith's rule is still used today by Scout groups and other casual hikers due to the ease of calculating walking time by hand using a paper map. However, despite the widespread use, Naismith's rule does have a well-known limitation; namely that the predicted speed does not change when descending a hill, regardless of the gradient. 

An alternative hiking function proposed by Tobler \cite{Tobler1993ThreeModelling}, has become more popular in recent research and other situations where speeds do not need to be calculated by hand:
 
\begin{equation*}
        W = 6*exp(-3.5|S + 0.05|),
\end{equation*}
where
\begin{quote}
    W = velocity (km/h)\\
    S = gradient of slope.
\end{quote}

\begin{figure}[!h]
    \includegraphics[width=\textwidth]{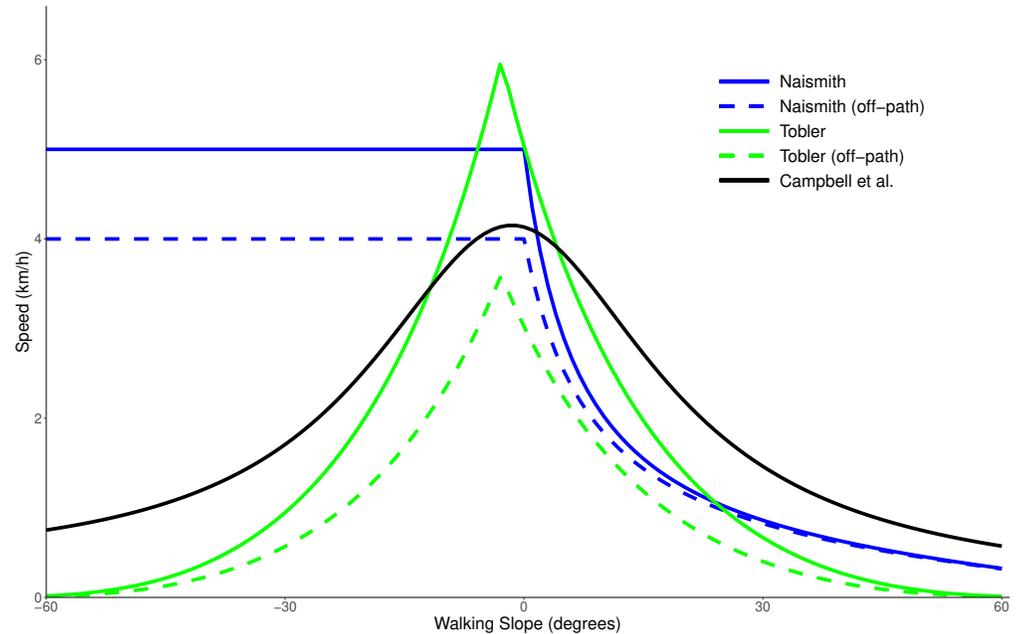}
    \caption{{\bf Existing functions used to calculate walking speed.} Naismith's rule \cite{Naismith1892CruachMore}, Tobler's hiking function \cite{Tobler1993ThreeModelling} and Campbell et al.'s function \cite{Campbell2022PredictingData} plotted as predicted walking speed in km/h against the slope in the direction of travel (walking slope) in degrees where positive is uphill. For Naismith's function and Tobler's function, on and off-path versions are shown.}
    \label{Fig1}
\end{figure}

Like Naismith's rule, this gives a speed of 5 km/h on flat ground, with a maximum speed of 6 km/h on a mild descent (around 3 degrees). In a similar manner to Aitken's correction, a factor of 0.6 is applied to the calculated speed for all off-road travel. Tobler's function avoids the issues seen in Naismith's rule when descending slopes, but it predicts a sharp peak in walking speed on mild descents, which may be unrealistic. The formulae discussed here are directly compared in Fig \ref{Fig1}.

Other studies have also looked at providing alternative methods to calculate walking speeds \cite{Irmischer2018MeasuringNavigation, Rees2004Least-costTerrain, Davey1994RunningApplications}, but all continue to use walking slope as the main variable to determine walking speed (with various multiplicative factors applied for off-road travel). 

When exploring speeds of fell-runners, Arnet \cite{Arnet2009ArithmeticalJapan} suggested that movement velocity may be dependent on three factors: obstruction (with different factors applied depending on the kind of obstruction), ascent in the run direction (walking slope) and slope of the terrain (hill slope). The actual values used in Arnet's calculations cannot be directly applied to walking speeds as they were based on orienteering championships where participants were running.

Experience tells us that traversing on a steep hill (while maintaining constant elevation) is more difficult than traversing flat ground. However, the existing methods estimate the same walking speed for both situations. Similarly, high levels of terrain obstruction in off-road areas (such as a thick gorse bush) are much more difficult to walk through than empty fields. The simple multipliers for off-road travel in Aitken’s correction and Tobler’s function do not provide any further distinction between two such regions. 

Wood and Schmidtlein \cite{Wood2012AnisotropicNorthwest}, took all three of Arnet's factors into account, and looked at evacuating citizens in the event of a hurricane. They applied Tobler's function to both the hill slopes and walking slopes, and calculated the terrain obstruction coefficients based on energy usage rather than walking speed (using \cite{Soule1972TerrainPrediction.}). They accepted that these were likely not the correct values, but were unable to find any better alternatives. Campbell, Dennison, and Butler \cite{Campbell2017AMapping} conducted a study using lidar data to explore the effects of ground roughness and vegetation density on firefighter evacuation speeds, but they did not consider the hill slope separately.

All of the studies mentioned above utilised relatively small sample sizes. However, the rise in use of global navigation satellite systems (GNSS), more frequently referred to as GPS tracking, means that a data-driven approach to modelling walking speed is now possible, which provides two main benefits. Firstly, it is possible to access GPS tracks from a wide variety of regions and terrains. Secondly, each track can easily be broken down into individual sections, enabling specific route features to be investigated at much higher spatio-temporal resolution. This has been explored in recent work \cite{Campbell2019UsingRates, Campbell2022PredictingData}, however the crowdsourced nature of these studies meant that data collection was not controlled, and thus that the data could not be assumed to consist wholly of walking or hiking tracks. In \cite{Campbell2019UsingRates}, data from hikes, jogs and runs was processed together, resulting in a very wide range of movement speed estimates. Campbell et al. attempted to overcome this in \cite{Campbell2022PredictingData} by only considering data points with a speed between 0.2 m/s and 5 m/s  (and the resulting model is shown in Fig \ref{Fig1}). However, 5 m/s (18 km/h) is much higher than the maximum predicted speeds from existing methods (such as Naismith's rule), so it is likely some non-walking data remained. Furthermore, applying a blanket 0.2 m/s minimum speed may well overlook valid datapoints recorded by particularly slow individuals, or in especially difficult regions. Finally, although these studies had the benefit of using large sample sizes, they both looked solely at the effect of the walking slope on speed, and did not explore additional variables. 

Here we used a data-driven approach to explore the impact of all three factors discussed by Arnet on walking speeds. These are the walking slope, the hill slope and the terrain obstruction. We aimed to use these factors to develop a model for the walking speed for an average individual. As with the existing methods, this model did not seek to model individual effects, and would still require tuning based on personal ability or conditions.

\section*{Materials and methods}

\subsection*{Data set, cleaning and key assumptions}
\label{2Data set cleaning and key assumptions}
Full details of the various datasets used in this study are provided in \nameref{S1_Appendix}. Further, a detailed description of the data filtering processes, and choices/assumptions made during data processing are described in \nameref{S2_Appendix}.

In summary, GPS tracks were obtained for hikes in the UK from Hikr.org \cite{Hikr.org2021UnitedReports} and OpenStreetMap (OSM) \cite{OpenStreetMap.org2021Tracks}. Elevation and walking slope values were calculated and added to every GPS point using data from the Ordnance Survey Terrain 5 Digital Terrain Map (DTM), which provides elevation data at 5 m intervals across the whole of the UK \cite{OrdnanceSurvey2020Terrain5}. Hill slope values were found using the quadratic surface method \cite{Zevenbergen1987QuantitativeTopography, Dunn1998TheGIS}. Each data point was then classified as on a paved road, on an unpaved road, or off road, determined by searching a 50 m radius around each point in an OSM Road dataset \cite{OpenStreetMap.org2021Data}. Paved and unpaved road classification was determined using \cite{OpenStreetMapWikiRoads}, with the unpaved road values being `path', `bridleway' and `track'.

Terrain obstruction information was calculated using lidar datasets \cite{LIDARDSMEngland, LIDARDTMEngland, LidarWales}, as the difference in values between a Digital Surface Map (DSM) and Digital Terrain Map (DTM). This meant that any physical feature which protruded from the ground was regarded as an obstruction. We had access to lidar data at 2 m resolution covering large areas of England and Wales, but the coverage was not complete. Of our off-road data ($\sim$2,900 km, spread across over 1,200 tracks), over 2,000 km had lidar data available. Exploration of the lidar data (see \nameref{S5_Appendix}) showed that there was a clear drop in walking speeds once the height of an obstruction was greater than 10 cm, beyond which the speed was relatively constant. We used this information to classify points into heavy obstruction (\textgreater 10 cm) or light obstruction ($<= 10$ cm) for modelling purposes.

\begin{figure}[!h]
    \begin{adjustwidth}{-1.75in}{0in} 
\includegraphics{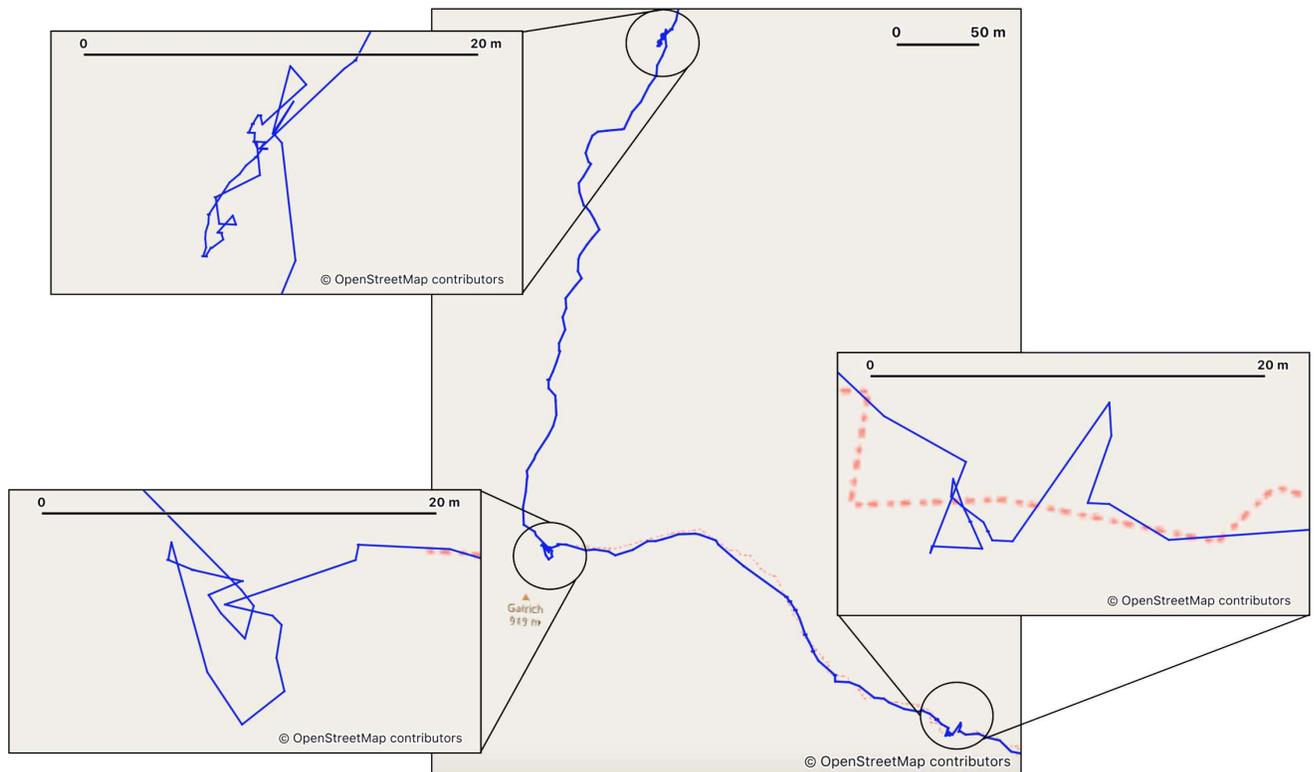}
\captionsetup{width=1\linewidth}
    \caption[width=\textwidth]{{\bf A GPS track where 3 breaks can be identified by finding point clusters.} Clusters of points can form on a GPS track when a break is taken during a hike. By identifying these clusters as potential breaks we are able to remove most break periods from the datasets used for our analysis of walking speeds. For full details of these and other data filtering methods see \nameref{S2_Appendix}. Background images from OpenStreetMap and OpenStreetMap Foundation \cite{OpenStreetMapMAPS}, visualised using QGIS \cite{OpenSourceGeospatialFoundationProjectQGIS.org2020}.}
    \label{Fig2}
    \end{adjustwidth}
\end{figure}

Visual inspection of the tracks showed that a large number contained long breaks which could impact the accuracy of a walking speed model. Fig \ref{Fig2} shows examples of regions where breaks are visible in a GPS track, and the process developed to identify these regions is outlined in Algorithm \ref{BreakfindingROUKAlg}.

\begin{algorithm}
\caption{Breakfinding process for a GPX track segment}
\label{BreakfindingROUKAlg}
\begin{algorithmic}[1]
\State Breakpoint\_list $ = \emptyset$
\State Find the median distance ($r_{median}$) and speed ($s_{median}$) of the segment
\For{point ($p_{i}$) in segment}
    \State Calculate travel direction quadrant and point angle
    \State Calculate break likelihood using the point speed and angle
    \If{speed == 0 or distance \textgreater 1 km or duration \textgreater 3 minutes}
        \State Breakpoint\_list += $p_{i}$
    \EndIf
    \If{speed \textgreater 10 km/h and duration($p_{i-1}$) \textgreater 3 minutes}
         \State Breakpoint\_list += $p_{i}$
    \EndIf
\EndFor
\For {point ($p$) in segment}
    \If{Neighbourhood of $p$ is a cluster ($C$)} \Comment{See Defs 1 \& 2, \nameref{S2_Appendix}}
        \For{point ($p_{c}$) in $C$}
            \If{Neighbourhood of $p_{c}$ is a new cluster ($C_{n}$)}
                \State $C = C \cap C_{n}$
            \EndIf
        \EndFor
        \State Remove points at the ends of the cluster with low break likelihood
        \State Add ‘missing’ points to the cluster (to make a continuous run of points) to form a Potential Break (B*)
        \If{less than half the points in B* have low break likelihood and there is travel in opposite quadrants (Q1 \& 3 or Q2 \& 4)}
             \State{Breakpoint\_list += B*}
        \EndIf
    \EndIf
\EndFor
\end{algorithmic}
\end{algorithm}

Where the datapoints in the original GPS track were under 50 m in length, they were merged together to minimise the effects of errors in the GPS location values. While doing this, the resulting distance was the sum of all distances in the constituent GPS points, so may be longer than the straight line distance between co-ordinates. Similarly, both hill and walking slope values, as well as obstruction height, were calculated as the weighted average of constituent points, weighted by point duration.

While the Hikr dataset consisted of tracks which were tagged as a walk or hike, within some of these there were segments where it was clear that the participant was driving to or from the hike location, based on the observed speeds. The OSM data, on the other hand, was not filtered by transport type. There were a large number of tracks which were clearly from faster modes of transport, as their speed was implausible for a hiker. A process to remove these non-walking tracks and segments was created, whereby the known Hikr walking segments were used to create filtering bounds of plausible walking speeds, which could then be applied to the remainder of the dataset. This process is summarised in Algorithm \ref{FilteringROUKAlg}.

\begin{algorithm}
\caption{Filtering process for GPS data from Hikr and OpenStreetMap}
\label{FilteringROUKAlg}
\begin{algorithmic}[1]
    \State Remove duplicate segments (containing sections with identical start location, end location, start time and duration)
    \State Remove all segments with median speed \textgreater 10 km/h
    \State Remove all breaks with duration \textgreater 30 seconds
    \State Remove all breaks containing points with speed \textgreater 10 km/h or distance \textgreater 1 km
    \State Merge remaining points into sections at least 50 m in length.
    \State Recursively remove points with speed \textgreater 10 km/h adjacent to a break, or the end of the track

\\
    \If{Hikr data}
        \If{segment mean speed \textgreater 10 km/h}
            \State remove segment
        \EndIf
        \State Calculate filtering bounds \Comment{Eq (\ref{Q1Hikrmax}) - (\ref{minHikrQ1}), \nameref{S2_Appendix}}
    \Else
        \State Identify Key Points \Comment{see \nameref{S2_Appendix}}
        \State Remove single datapoints between Key Points
        \State Remove points where median speed between consecutive key points \textgreater Eq (\ref{Q1Hikrmax})
        \While{segment length is not consistent}
            \State Remove points with speed \textgreater 10 km/h adjacent to a break, or the end of the track
            \If{ segment median speed \textgreater Eq (\ref{Q1Hikrmax}) \textbf{or}
            segment minimum speed \textgreater Eq (\ref{medHikrmed}) \textbf{or}
            segment upper quartile speed \textgreater Eq (\ref{topHikrmax}) \textbf{or}
            segment upper whisker speed \textless Eq (\ref{minHikrQ1}) \textbf{or} segment duretion \textless 2.5 minutes
            }
                \State Remove segment
            \EndIf
        \EndWhile
    \EndIf
    \\
    \State Combine all segments into a single dataset
    \State Remove the fastest and slowest 0.5\% of the data 
\end{algorithmic}
\end{algorithm}

Following this, a decision was made to remove data from tracks found in Scotland. Lidar data covering the walking tracks was necessary to model the terrain obstruction, and was not sufficiently available in Scotland at the time of the study. Furthermore, analysis showed that that walking speeds in Scotland were at the extreme end of what is seen throughout the rest of the UK (see \nameref{S4_Appendix}). Including this data without also including a corresponding extreme dataset where lidar data is available may result in incorrect modelling. All OSM track segments which took place within Scotland were excluded from further processing. Similarly Hikr tracks which were tagged as taking place in Scotland, and which fully took place in Scotland were excluded.

Our final modelling dataset consisted of 7,636 GPS tracks from England and Wales, with over 1.4 million individual data points and almost 88,000 km of travel. Each datapoint represented approximately 50-100 m of travel, and contained:
\begin{itemize}
    \setlength\itemsep{0em}
    \item Start coordinate
    \item End coordinate
    \item Start time
    \item Duration
    \item Distance
    \item Speed
    \item Elevation
    \item Walking slope
    \item Hill slope
    \item On-road flag
    \item Paved road flag (if on-road)
    \item Obstruction data available flag (if off-road)
    \item Heavy obstruction flag (if off-road and obstruction data available)
\end{itemize}

\subsection*{Modelling}

\subsubsection*{Model Formulation}

Pilot studies were conducted to identify an appropriate model framework, using tracks within Scotland (see \nameref{S3_Appendix}). Generalised linear model (GLM) and generalised additive model (GAM) approaches were explored, and within both we looked at the relationship between the walking and hill slopes, and the walking speed, with a small number of prior assumptions. As it is more challenging to walk on steeper slopes, for both the hill and walking slope components we knew that the walking speed should be a decreasing function of the magnitude of slope (with some allowance for faster walking speeds on mild descents). Models which failed to predict this were removed under the assumption that the data were overfitted. Furthermore, previous work \cite{Davey1994RunningApplications,Kay2012RouteTerrain,Balstrm2002OnTerrain,Llobera2007Zigzagging:Strategies} has identified the existence of a critical gradient; the angle at which it is faster to zig-zag up a hill, rather than ascend directly. This occurs at a walking slope of around 15 -- 21 degrees, so models which failed to predict the critical gradient occurring below 21 degrees were removed.

10-fold cross-validation was used to compare the remaining model parameters, looking at R-squared values, root-mean-squared error (RMSE) and mean absolute error. Where multiple models performed equally well, the simplest model was selected for ease of interpretabilty and real-world application. The selected model type was a Generalised Linear Model (GLM). Models were implemented using R version 3.6.1 \cite{R}. 

\subsubsection*{Terrain Types}

Each of the three road types (paved road, unpaved road, off-road) was included in the model, both as factor variables, and as interaction terms with each of the slope variables.

Before adding terrain obstruction data to the model, we checked that there was no systematic difference between the walking speeds in regions where we had lidar data, and regions where we did not (see \nameref{S5_Appendix}). Thus our findings in regions where lidar data was available could be extended to those where it was unavailable. Factor variables were then added to the model for each obstruction level (heavy, light or unknown obstruction).

\subsubsection*{Statistical Analysis}

Variables within the model were tested for significance using the Wald test, which allows us to account for correlation between points within the same track (\texttt{coeftest} function within \texttt{lmtest} package in R).

To measure the impact of our model, we compared walking speed predictions of our model against those of Naismith's, Tobler's and Campbell et al.'s models. Four different metrics were compared; the average percentage error, mean squared error (MSE), root-mean squared error (RMSE) and R squared value. These were explored when looking at both individual 50 m track sections, as well as predicted walking times for tracks as a whole. Finally, we isolated the off-road track sections in order to assess the improvement of our model at predicting walking speeds for off-path travel.

\subsection*{Code Availability}
\label{0Code}

Documented code written for this study is available online in a Github repository \texttt{\href{https://github.com/AndrewWood94/PhDThesis}{AndrewWood94/PhDThesis}}, and is licensed under the terms of the GNU General Public License v3.0.
% Results and Discussion can be combined.
\section*{Results}
We started by assembling a dataset of hikes derived from approximately 20,000 public GPS tracks. These tracks recorded a variety of transport methods and required significant filtering. This process included iterative data cleaning to remove erroneous or non-walking data and identify/remove breaks (e.g. Fig \ref{Fig2}) to give us a final usable dataset containing 7,636 GPS tracks, with over 1.4 million individual data points and covering almost 88,000 km of travel in the U.K. Each data point represents at least 50 m of travel (with a mean distance of 60.3 m), and the breakdown of the data by slope angle and terrain type is shown in Table \ref{tab:1Terrain data distances}. Previous research has found that most walking takes place on low walking slopes \cite{Proffitt1995PerceivingSlant}, and this is evidenced by our data ($\sim$98\% of our data was from walking slopes of under 10 degrees).

\begin{table}[!ht]
\begin{adjustwidth}{-0.65in}{0in}
    \centering
    \caption{Total distance of data under different terrain conditions (km)}
    \begin{tabular}{|l+c|c|c+c|c|c|}
    \hline
    & \multicolumn{3}{|l+}{\bf Hill Slope (degrees)} & \multicolumn{3}{|l|}{$\mathbf{\vert Walking \  Slope \vert 
 \ (degrees)}$}\\
    \hline
    & 0-10 & 10-20 & \textgreater{}20 & 0-10 & 10-20 & \textgreater{}20 \\
    \thickhline
    Paved road & 62159.1 & 7841.2 & 2081.9 & 70726.5 & 1277.3 & 78.4 \\ 
    \hline
    Unpaved road & 9996.9 & 2210.3 & 700.7 & 12421.7 & 460.0 & 26.2 \\
    \hline
    Off Road (obstruction unknown) & 773.5 & 114.2 & 17.8 & 871.7 & 31.7 & 2.0 \\
    \hline
    Off Road (light obstruction) & 1282.9 & 150.1 & 23.8 & 1424.6 & 30.6 & 1.7 \\ 
    \hline
    Off Road (heavy obstruction) & 428.7 & 105.2 & 28.5 & 543.5 & 18.5 & 0.4 \\ 
    \hline
    \end{tabular}
    \label{tab:1Terrain data distances}
\end{adjustwidth}
\end{table}

Our curated hike dataset allowed us to create a data-driven model which we can directly compare with existing walking speed algorithms. The model formulation was selected using a small-scale exploratory study which considered data from Scotland (see \nameref{S3_Appendix}). In this exploratory study, multiple different model types were explored which could fit the data, and which matched existing knowledge about walking speeds. Cross-validation methods showed that there was very little difference in performance of the best models, therefore the final model was a Generalised Linear Model (GLM), which was chosen as it was the simplest of those tested (we had no evidence that a more complex model would be superior). This choice also meant that our model was both easy to interpret, and simple to apply to future work.

This final GLM model included all three of the variables suggested by Arnet \cite{Arnet2009ArithmeticalJapan}:

\begin{equation}
    v = exp(a+b\phi+c\theta+d\theta^2)
\end{equation}
where
\begin{quote}
$v = \text{walking speed (km/h)}$\\
$\phi = \text{hill slope angle (degrees)}$\\
$\theta = \text{walking slope angle (degrees)}$
\end{quote}

Terrain obstruction level was included as a factor variable, while we considered the road types as both factor variables and interaction terms. Not all terms had a significant effect on all variables; we therefore created a model with all possible terms, and removed them one at a time (in order of least significance) until all remaining terms were significant to at least 95\% confidence  level (using Wald test). The final values for a, b, c and d are given in Table \ref{tab:2ROUK model variable values} for each of the terrain obstruction levels and road types. The critical gradient for this model is between 14 -- 16 degrees when walking uphill and -16 -- -18 degrees when walking downhill (depending on road and obstruction conditions), which is in line with previous findings. 

\begin{table}[!ht]
\begin{adjustwidth}{-0.5in}{0in}
    \centering
    \caption{Final walking speed model variable coefficients}
    \begin{tabular}{|l+c|c|c|c|}
    \hline
    & $a$ & $b$ & $c$  & $d$ \\ 
    \thickhline
    Paved road & 1.580 & -0.00389 & -0.00726 & -0.00218 \\ 
    \hline
    Unpaved road & 1.580 & -0.00389 & -0.00965 & -0.00248 \\
    \hline
    Off-road (obstruction unknown) & 1.536 & -0.00731 & -0.00965 & -0.00187 \\
    \hline
    Off-road (light obstruction) & 1.580 & -0.00731 & -0.00965 & -0.00187 \\ 
    \hline
    Off-road (heavy obstruction) & 1.443 & -0.00731 & -0.00965 & -0.00187 \\ 
    \hline
    \end{tabular}
    \label{tab:2ROUK model variable values}
\end{adjustwidth}
\end{table}

\begin{figure}[!h]
\begin{adjustwidth}{-2.25in}{0in} 
    \includegraphics[width=\linewidth]{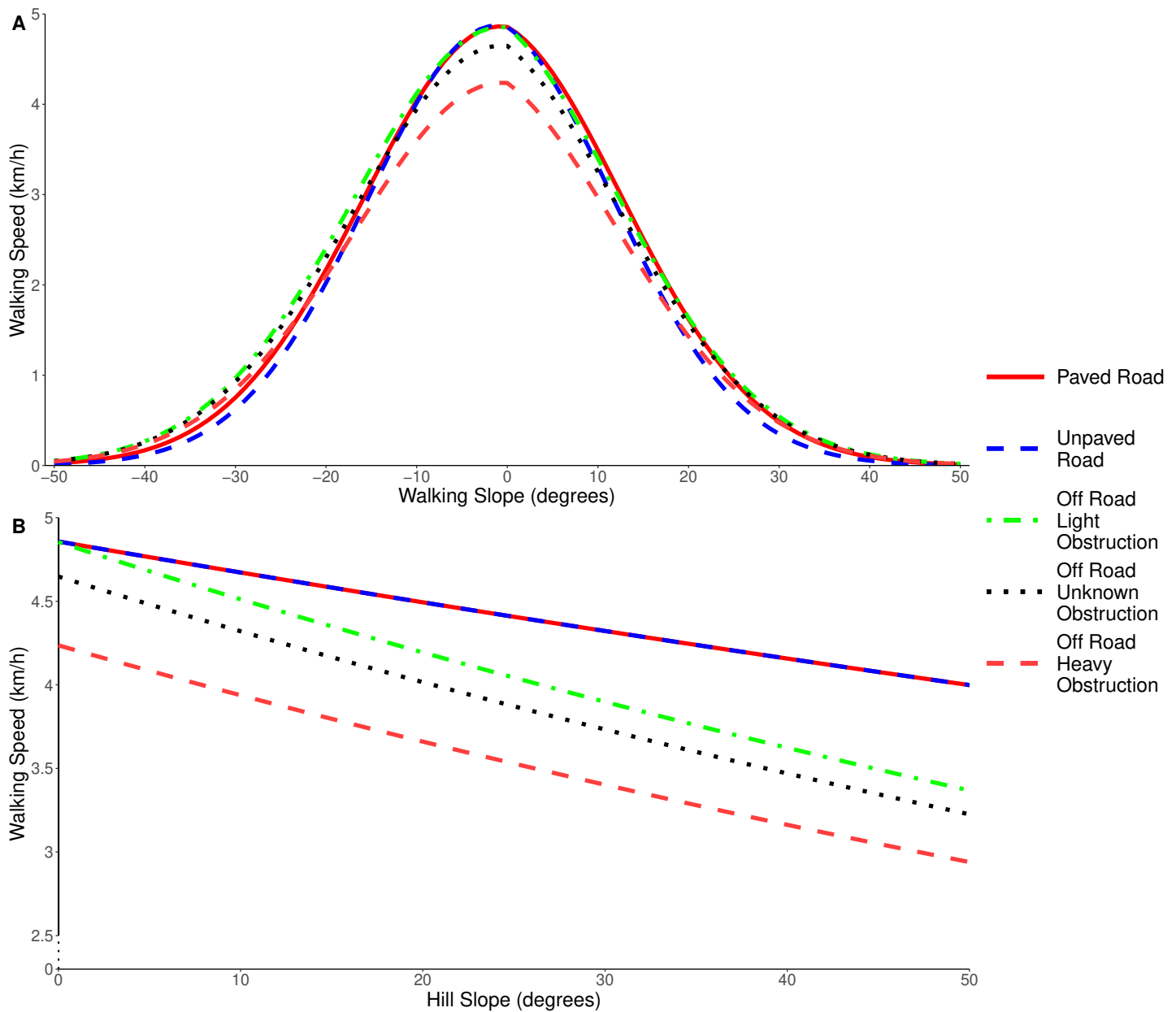}
    \captionsetup{width=1\linewidth}
    \caption[width=\textwidth]{{\bf Walking speed predictions under different terrain conditions.}  When: (A) travelling directly up or down hills of varying slope (walking slope), (B) traversing across hills of varying slope (hill slope).}
    \label{Fig3}
    \end{adjustwidth}
\end{figure}

Fig \ref{Fig3} shows the predicted walking speeds under different conditions. The importance of including both the hill slope and terrain obstruction variables can be clearly seen when looking at the Off Road Light Obstruction speed predictions. When directly ascending or descending a slope, the walking speed is comparable to walking on a road. However, when traversing a slope while off road, the walking speed is comparable to traversing a slope of double the gradient while on a road or path. Similarly, comparing the walking speed predictions of Off Road Light Obstruction and Off Road Heavy Obstruction reveals that just 10 cm of vegetation (our cutoff point for heavy obstruction) can reduce the walking speed by more than 0.5 km/h.

\begin{figure}[!h]
\begin{adjustwidth}{-2.25in}{0in} 
    \includegraphics[width=\linewidth]{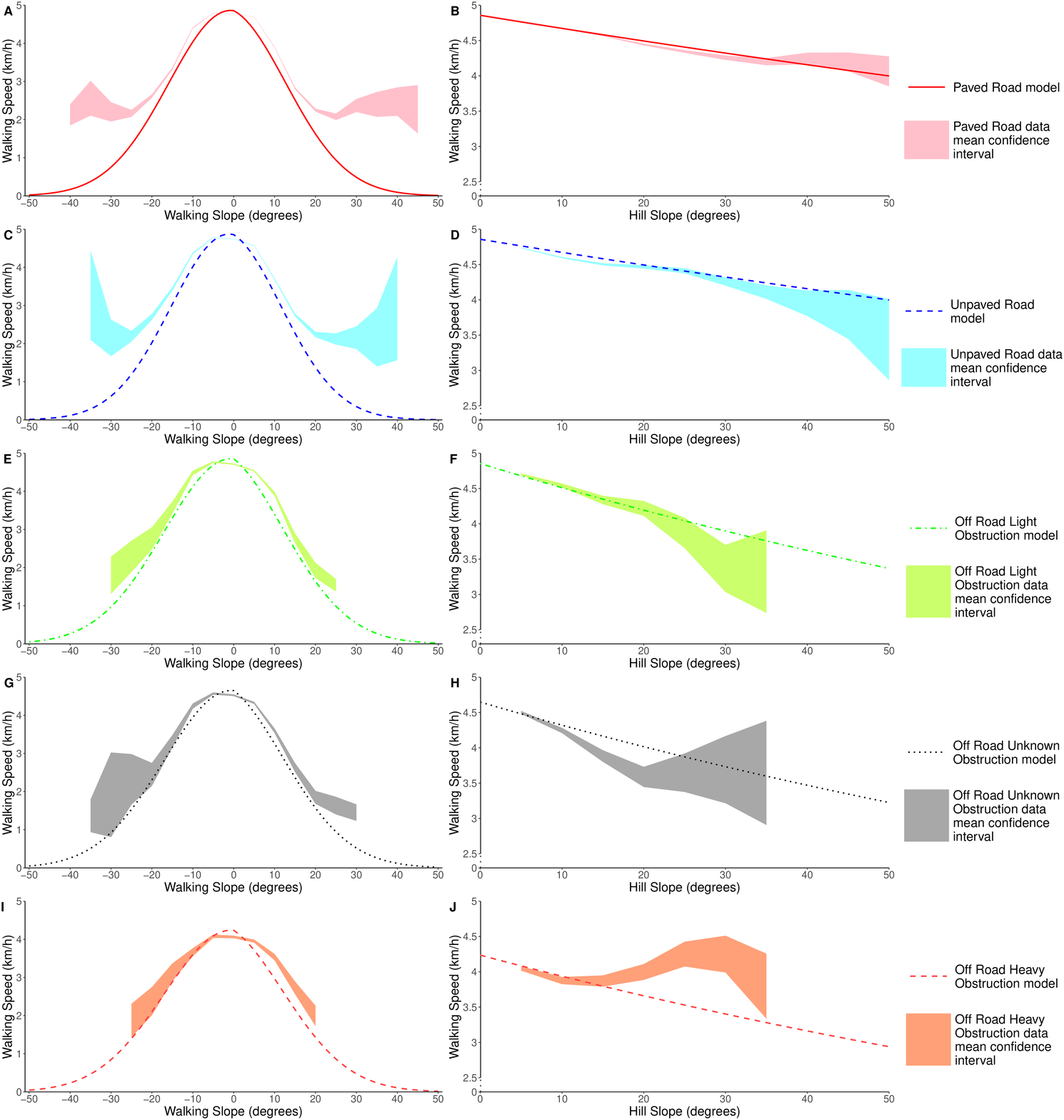}
    \captionsetup{width=1\linewidth}
    \caption[width=\textwidth]{{\bf Walking speed predictions under different terrain conditions.}  When: (A,C,E,G,I) travelling directly up or down hills of varying slope (walking slope), (B,D,F,H,J) traversing across hills of varying slope (hill slope). Also shown in each plot is the 95\% confidence interval of the mean value of the walking speed for the terrain type, calculated at 5 degree intervals, using data bins with a width of 10 degrees. Note that the confidence intervals were calculated using only data which is within 5 degrees of directly ascending (A,C,E,G,I) or traversing (B,D,F,H,J) the slope.}
    \label{Fig4}
    \end{adjustwidth}
\end{figure}

Fig \ref{Fig4} shows the same walking speed predictions as Fig \ref{Fig3}, alongside the confidence interval for the mean walking speed for each terrain type. In the low-slope regions where most walking occurs, our model fits closely with the mean data confidence intervals. Our model does deviate from the confidence interval in some areas, particularly in high-slope and off-road regions. However, these are also the areas where we have the least amount of data (see Table \ref{tab:1Terrain data distances}). In Fig \ref{Fig4}J the confidence interval for the mean would suggest that it is faster to walk on hill slopes of 30 degrees than hill slopes of 10 degrees. We have less than 30 km of data recorded in heavy obstruction regions on hill slopes of over 20 degrees, and less than 20 km of this had a walking slope magnitude of under 5 degrees (indicating that the slope was being traversed). Further, even within this range, the data is skewed towards the lower hill slope values. This lack of data explains the widening confidence interval, and counter-intuitive observations and we suggest that a targeted study would be required to collect more data in this region.

\begin{figure}[!h]
\begin{adjustwidth}{-2.25in}{0in} 
    \includegraphics[width=\linewidth]{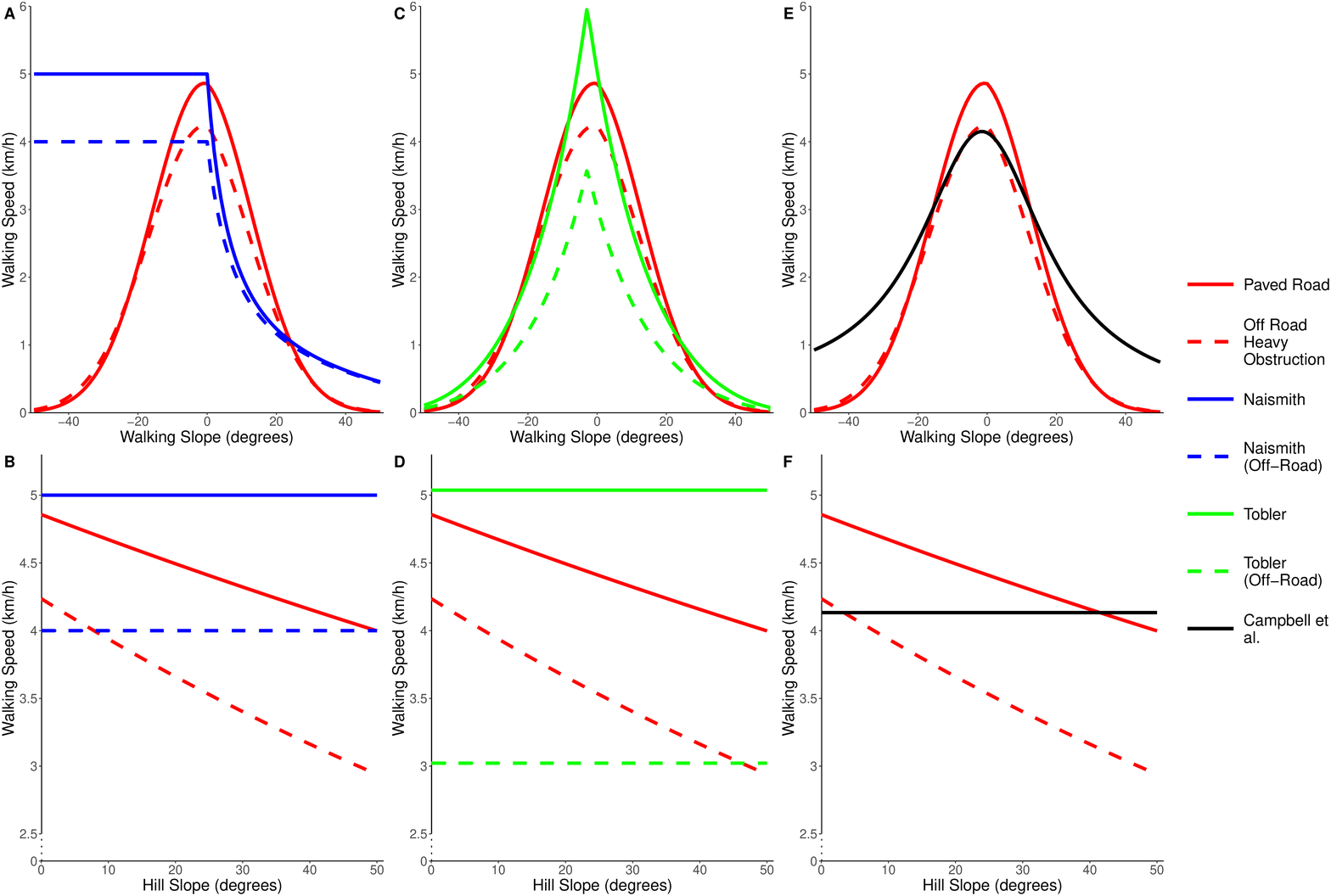}
    \captionsetup{width=1\linewidth}
    \caption[width=\textwidth]{{\bf Comparison of new model and existing hiking functions.}  Predicted walking speeds of the new model, Naismith's rule, Tobler's function and Campbell et al.'s function when: (A, C, E) travelling directly up or down hills of varying slope (walking slope), (B, D, F) traversing across hills of varying slope (hill slope).}
    \label{Fig5}
\end{adjustwidth}
\end{figure}

Fig \ref{Fig5} compares the Paved Road and Off Road Heavy Obstruction speed predictions from our model against the existing functions from Naismith, Tobler and Campbell et al. When looking at the walking slope, the largest areas of deviation between our model and Naismith's rule occurs when descending a slope, as Naismith's rule does not predict a reduced speed in this scenario. For both Tobler's and Campbell et al.'s functions, the shape of the walking slope component is relatively similar to our new model, with the main distinction being the peak predicted speed on flat ground. None of the existing functions account for the hill slope, which leads to large disparities when predicting the walking speed for slope traversals. A further example of this can be seen in \nameref{S6_Appendix}, which shows the walking speeds for a simulated off-road route which encounters the full range of hill and walking slopes.

When comparing the performances of each of the models (Table \ref{tab:3comparison}), the predicted speeds for individual 50 m sections had a lower RMSE and percentage error, and a higher R squared value using our new model than in the existing ones. The R-squared value is still very low, however we suggest that this is due to the variability within the data. We have previously acknowledged that there are many individual effects which can impact the walking speed, and which we did not attempt to capture in our model. Instead it captures the general trend of the walking speed for an average individual under average conditions, and does this better than existing models (evidenced by the improved RMSE).

\begin{table}[!ht]
\centering
\caption{Comparison of new model against existing methods to calculate walking speeds.}
\begin{tabular}{|l|c|c|c|c|}
\hline
& New Model & Naismith & Tobler & Campbell\\
\hline
Average \% error & 23.68 & 26.36 & 26.17 & 25.33\\
\hline
MSE & 1.20 & 1.61 & 1.53 & 1.58\\
\hline
RMSE & 1.10 & 1.27 & 1.24 & 1.26\\
\hline
R\textsuperscript{2}  & 0.09 & -0.22 & -0.16 & -0.19\\
\hline
\end{tabular}
\label{tab:3comparison}  
\end{table}

\begin{figure}[!h]    
\begin{adjustwidth}{-2.25in}{0in} 
    \includegraphics[width=\linewidth]{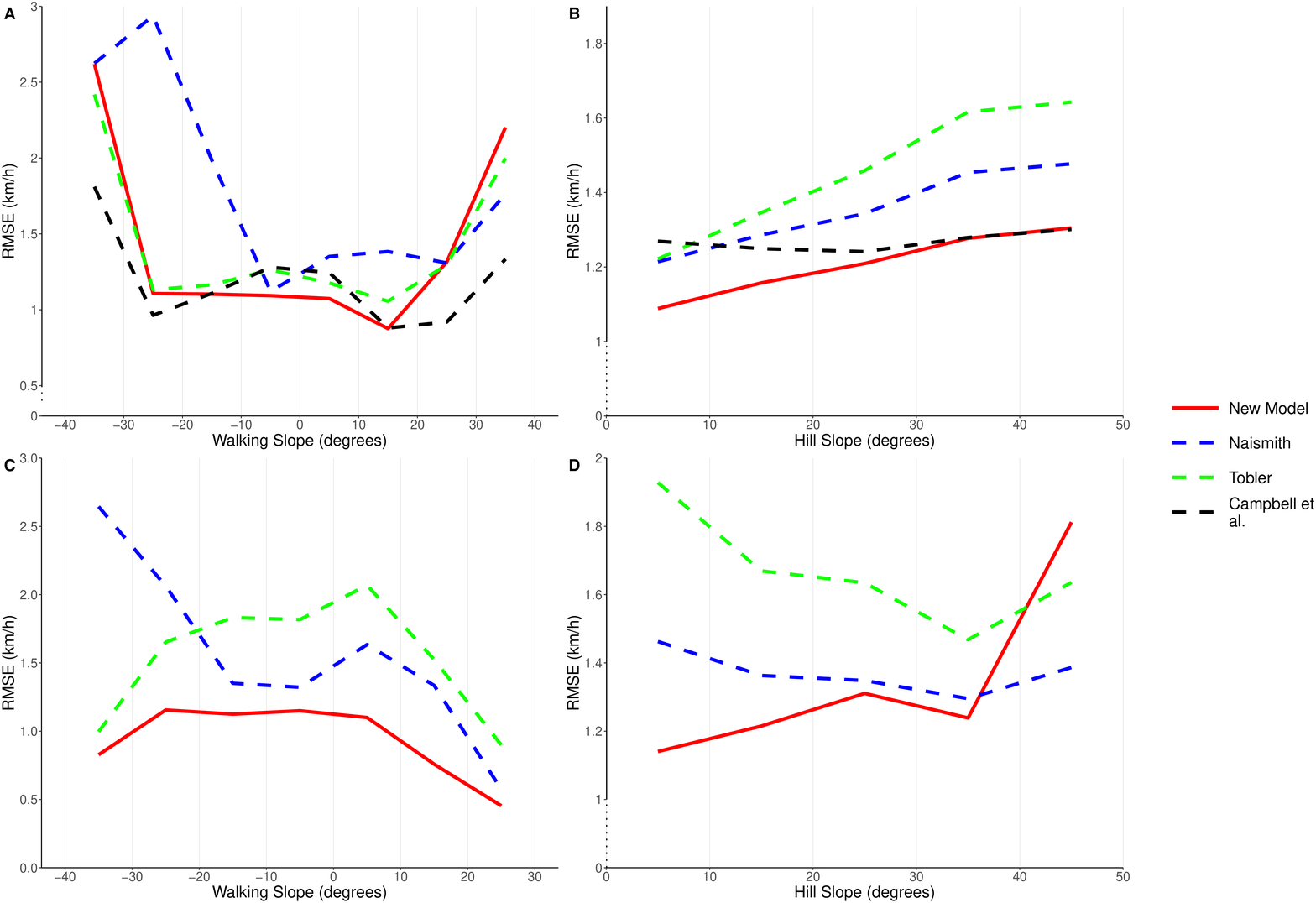}
    \captionsetup{width=1\linewidth}
    \caption[width=\textwidth]{{\bf Comparing RMSE values for the new model, Naismith's rule, Tobler's function and Campbell et al.'s function.} When: (A) travelling directly up or down hills of varying slope (all data, walking slope), (B) traversing across hills of varying slope (all data, hill slope), (C) travelling directly up or down hills of varying slope (off-road data only, walking slope), (D) traversing across hills of varying slope (off-road data only, hill slope). Campbell et al.'s function does not provide off-road speed estimates, so was not included in the off-road data comparisons.}
    \label{Fig6}
\end{adjustwidth}
\end{figure}

\begin{figure}[!h]
    \begin{adjustwidth}{-2.25in}{0in} 
    \includegraphics[width=\linewidth]{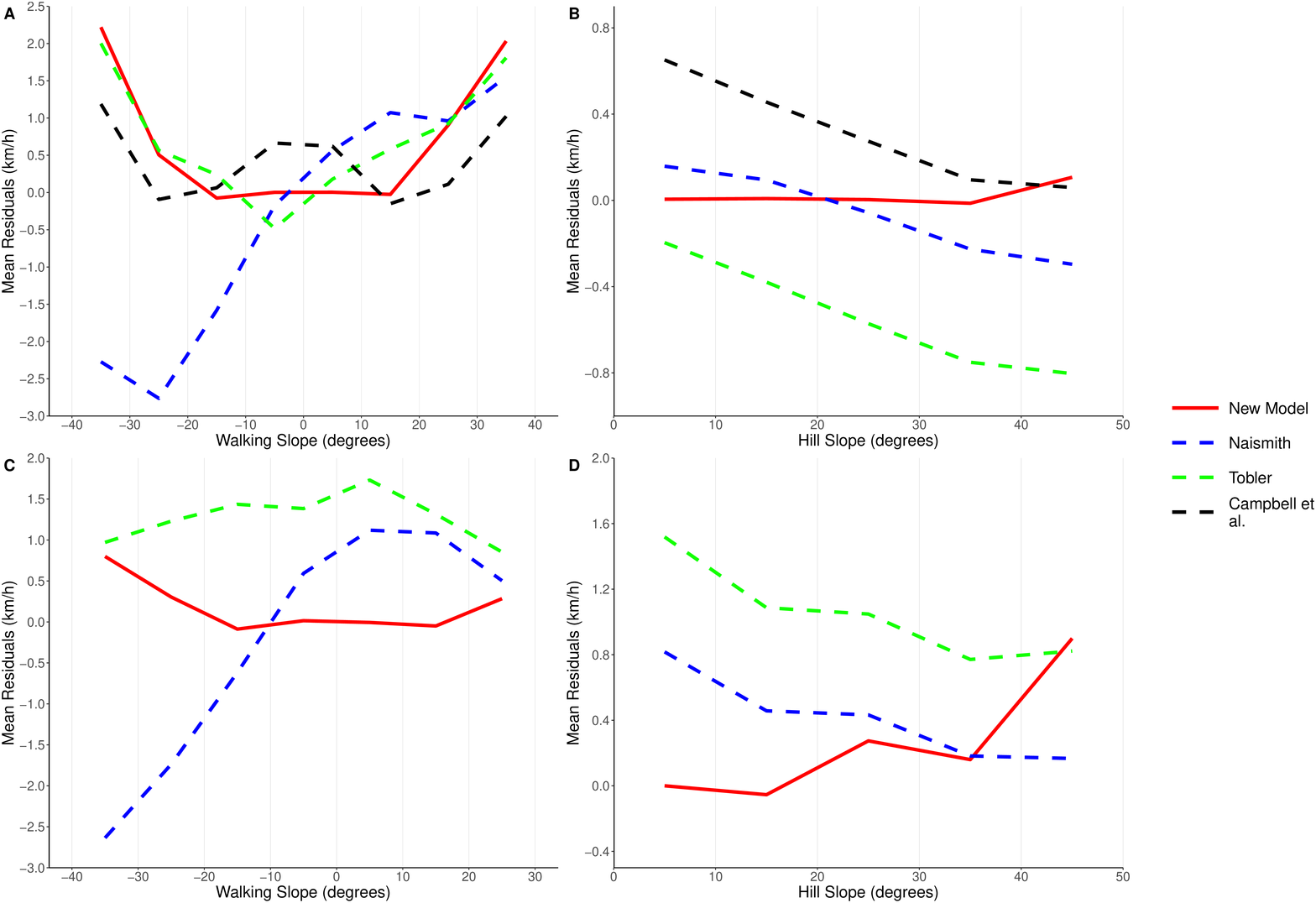}
    \captionsetup{width=1\linewidth}
    \caption[width=\textwidth]{{\bf Comparing mean residual values for the new model, Naismith's rule, Tobler's function and Campbell et al.'s function.} When: (A) travelling directly up or down hills of varying slope (all data, walking slope), (B) traversing across hills of varying slope (all data, hill slope), (C) travelling directly up or down hills of varying slope (off-road data only, walking slope), (D) traversing across hills of varying slope (off-road data only, hill slope). Campbell et al.'s function does not provide off-road speed estimates, so was not included in the off-road data comparisons.}
    \label{Fig7}
\end{adjustwidth}
\end{figure}

To isolate the impact of each of the slope variables, we filtered the results to look at the data where a slope was being directly climbed or traversed. Figs \ref{Fig6}A, B and \ref{Fig7}A, B show the RMSE and mean residuals for each of the models, for data which was within 5 degrees of directly climbing (A) or traversing (B) hills of varying slope. From this we can clearly see that Naismith's rule consistently overestimates walking speeds when descending a slope, and underestimates speeds when climbing a slope. When ascending or descending a slope, the RMSE of our GLM is similar to that of Tobler's hiking function. However, one of the main areas where we see an improvement using our model is on slight declines. Tobler's hiking function suggests that walking speed increases on mild descents up to a maximum of 6 km/h. It is clear from Fig \ref{Fig6}A, that Tobler's function overestimates the walking speed in this region. Campbell et al.'s function has a slightly lower RMSE value than our new model on the steepest walking slopes, however it underestimates the walking speeds on flat ground and mild slopes; the regions where most walking occurs. Improved walking speed predictions in this region therefore have the greatest impact in real-world situations. Within this region our model consistently has a lower RMSE than the existing functions, and a mean residual error close to 0 km/h. 

 We also see an improvement in RMSE when using our model to predict speeds for hill traversals (Fig \ref{Fig6}B). We can note from Fig \ref{Fig7}B that both Naismith's rule and Tobler's hiking function consistently overestimate the walking speed when traversing a slope, as they do not take into account the impact that the hill slope has on reducing walking speeds. The performance of Campbell et al's model improves as the hill slope increases, although we suggest this is more due to it underestimating the speed on shallow slopes. We do see that the average error in our model increases as the hill slope increases, but we believe that this is due to limited volumes of data at high hill slopes ($\sim$0.5\% of our data occurs on hill slopes steeper than 40 degrees). 

As well as looking at the overall performance of our new model, we looked to explore how well our model performed in off-road conditions, compared to the off-road adjustments for the existing functions (Naismith's reduced base speed of 4 km/h, and Tobler's correction factor of 0.6). Figs \ref{Fig6}C, D and \ref{Fig7}C, D show the RMSE and mean residuals, only considering data which was recorded in off-road conditions. From Figs \ref{Fig6}C and \ref{Fig7}C it is clear that Tobler's function consistently underestimates the walking speed when off-road. The factor of 0.6 is a larger reduction in walking speed than is observed in practice. As we found when looking at our data as a whole, Naismith's rule underestimates the walking speed when climbing a slope and overestimates when descending a slope. Our new model does not suffer from these problems, with both a lower RMSE and lower absolute mean residual value across all walking slopes. Both of these existing models also consistently underestimate walking speeds when traversing a slope, unlike our new model which has a mean residual of less than 0.4 km/h on slopes of up to 35 degrees. The error in predictions of our new model does increase as the hill slope increases, though the RMSE is generally lower than seen in the existing models. On the steepest hill slopes our model appears to perform less well than the existing ones, though only 0.2\% of our off-road data occurred on a hill slope steeper than 40 degrees. 

Although we have shown an improvement in walking speed predictions over short sections of routes, this did not translate to similar results when looking at predicted walking times for routes as a whole. Our model and all of the existing models which we have explored here had an average percentage error of 13.5\% - 15.5\% when predicting the time taken for a complete route. However, based on the errors seen in Figs \ref{Fig6} and \ref{Fig7}, we believe that this is a result of errors cancelling out over the course of a hike. For example while ascending a hill, Naismith's rule will underestimate the walking speed (and thus overestimate the walking time), but it will then overestimate the walking speed on the subsequent descent, leading to a relatively accurate total time estimate. The results here suggest that Naismith's rule, and other existing functions, are still a good rule of thumb to calculate route times as a whole, but time estimates for individual sections of a route will be less accurate than when using the new model found here.

\section*{Discussion}

We have developed a model for walking speed which is very robust, due the large volume of data (88,000 km) used to build it, and which correlates with the data over a wider range of conditions than commonly used formulae. Data from tracks confirms that each of the walking slope, the hill slope and the terrain type or obstruction are significant factors in determining walking speeds. The model improves on existing methods to predict walking speeds (Figs \ref{Fig6} \& \ref{Fig7}). We have also shown the specific improvement that our new model has on predicting walking speeds in off-road conditions, compared to the simple off-road speed reductions used by existing models. The existing methods to calculate walking speeds require tuning for use in real-world scenarios, as there are many factors which can affect an individual's walking speed beyond the slope and obstruction level (such as the weather, fitness level or age) \cite{Mountaineering2023Planning, Miao2023AnalysisEnvironment, Liang2020HowAreas}. The model presented here requires the same tuning as these existing methods but provides more a more accurate population average walking speed across a wide range of terrain and slope conditions.

Our results confirm that Naismith's rule (Fig \ref{Fig1}) is still a good rule-of-thumb to use when estimating the total walking time for a route, especially in situations where the calculation must be done by hand. However, the findings here can be used as an addition to Naismith's rule; it is likely that (under Naismith's rule) the predicted ascent time will be overestimated and the predicted descent time will be underestimated. It is not uncommon for hikers to contact one another when they reach the summit of a hill, and provide an estimated arrival time back at the campsite. Knowing that the descent will likely take longer than estimated by Naismith's rule will result in more accurate arrival estimations being given. Similarly, the knowledge of how the hill slope reduces walking speeds, or that just 10 cm of vegetation can reduce walking speeds by up to 0.6 km/h may well affect route choices made when out on a walk. For example, if a hiker is following a footpath, but can see from their map that the path forms a large curve then they can use our findings to decide whether it will be faster to travel off-road and cut the corner. On flat terrain with heavy levels of obstruction, our model suggests that such a short cut will be faster if the distance covered on the path is more than 15\% longer than the off-road distance. Speed is not the only factor which would affect this decision, as safety and navigability are also important variables, but these results can help people make more informed choices when on a hike.

The benefit of using crowdsourced GPS data to build our model is also a limitation of the approach, as we did not have control over data collection. This meant that models were unable to account for any bias in our data such as group size, ability and composition, or other potential variables such as weather conditions, as factors in determining walking speed (although we would expect the volume of data to cause most of these effects to average out). 

Unlike previous work \cite{Campbell2022PredictingData}, we did not use fixed values to classify breaks and non-walking or hiking tracks. Instead we developed filters based on the attributes of known walking data (see \nameref{S2_Appendix}). The methods used to filter the datasets were blinded to the outcome of the model generation, the choice of filtering methods will have had an impact on the dataset and subsequent model and no ground truth was available against which to test our assumptions. 

Our method of calculating the terrain obstruction value was relatively crude, looking only at the obstruction height at each GPS point. While this did prove to be successful, and we observed a clear difference in walking speeds between areas of light and heavy obstruction (see \nameref{S5_Appendix}), the inaccuracies present within GPS data may have led to some erroneous obstruction measurements, for example in a field sparsely populated with trees. In future, efforts should be made to refine this approach, such as considering the average obstruction level over a wider area around each point.

A further limitation of our data came when we looked to classify points into paved roads, unpaved roads or off-road. A combination of GPS drift and map error means that there is significant uncertainty and so we had to use a search radius around each data point to identify potential roads. We suspect that we were likely overclassifying tracks on roads. While our model appears to be robust to this overclassification (due to the volumes of correctly classified data used), the overclassification left us with a reduced number of off-road datapoints from which to predict off-road travel speeds.

Furthermore, the use of crowdsourced data meant that all of our data came from `walkable' regions by definition. When including the terrain obstruction variable, we were unable to determine if there are levels of terrain obstruction which makes walking impossible. Similarly, the vast majority of the data was collected on shallow hill- and walking slopes, leading to a sparcity of data in steeper areas. While this does mean that we can be very confident about our walking speed predictions in less steep regions (where most walking occurs), it is unclear whether the lack of data on steeper regions is a result of steep slopes being relatively rare, or that they cannot be easily navigated, so hikers chose an alternate path. 
As described above we had to make a number of assumptions regarding data filtering and processing including model selection, and other choices may give different results. To support anyone who wants to challenge or test these assumptions, or try different models, we have made all our code available on Github. Further, all of the data sources used are detailed in \nameref{S1_Appendix} and the filters/assumptions we used to clean the data are fully detailed in \nameref{S2_Appendix}.
\section*{Conclusion}

Widely used algorithms (e.g. Naismith's rule) for estimating walking/hiking speed are simple to understand, very easy to calculate but are based on limited observations. Here we curated a dataset of almost 88,000 km of walking and hiking data. We found that the existing algorithms perform quite well against the dataset but they tend to overestimate ascent time, underestimate descent time and most ignore terrain obstruction and hill slope both of which we found to be significant factors. We used the data to derive a new model that takes into account these variables. We demonstrated that the model provides more accurate walking speeds than the existing methods in all scenarios, and particularly in off-road regions. By providing improved walking speed predictions in these off-road regions, we have enabled more accurate calculations of the fastest route to or from any given location, which could save minutes in an emergency situation where every second is important.

% Include only the SI item label in the paragraph heading. Use the \nameref{label} command to cite SI items in the text.
\section*{Acknowledgments}

Preprocessing of the GPX files made use of the resources provided by the Edinburgh Compute and Data Facility (ECDF) \cite{Eddie2022website}.

\nolinenumbers

\bibliography{Bibliography}

\section*{Supporting information}

\paragraph*{S1 Supporting Information.}
\label{S1_Appendix}
{\bf Data sources table}

\paragraph*{S2 Supporting Information}
\label{S2_Appendix}
{\bf Data acquisition and preparation}

\paragraph*{S3 Supporting Information.}
\label{S3_Appendix}
{\bf Exploratory data modelling study.} 

\paragraph*{S4 Supporting Information.}
\label{S4_Appendix}
{\bf Exploring the differences between Scotland and the rest of the UK.} 

\paragraph*{S5 Supporting Information}
\label{S5_Appendix}
{\bf Exploring the impact of terrain obstruction.}

\paragraph*{S6 Supporting Information}
\label{S6_Appendix}
{\bf Comparison of walking speed changes while crossing a simulated off-road terrain region.}
\section*{S1 Supporting Information. Data sources table}

\begin{table}[!h]
\begin{adjustwidth}{-1in}{0in}
\centering
\caption{Summary of data sources used during this work}
\begin{tabular}{| L{0.17\linewidth} | L{0.21\linewidth} | C{0.15\linewidth} | L{0.35\linewidth}|}
\hline
\centering Data Type & \centering Data Source & Download Date & \centering Notes \tabularnewline
\thickhline 
\raggedright Hikr GPS data & Hikr.org & 01-07-2021 & Within the UK data\textsuperscript{1}, only tracks which took place within Scotland\textsuperscript{2} were used for exploratory study (see \nameref{S3_Appendix}).\\
\hline
OpenStreetMap GPS data & OpenStreetMap.org & 01-07-2021 & Accessed using planet.gpx
regional extracts\textsuperscript{3}\\
\hline
Ordnance Survey elevation data & Ordnace Survey Terrain 5 DTM & 05-08-2021 & Accessed using EDINA Digimap Ordnance Survey Service\textsuperscript{4}\\
\hline
OpenStreetMap road data & OpenStreetMap.org & 04-08-2021 & Accessed using planet.osm
regional extracts\textsuperscript{5}
\\ 
\hline
England lidar data & National LIDAR Programme & 16-09-2021 & 2m resolution data was used and accessed using EDINA LIDAR Digimap Service\textsuperscript{4}\\ 
\hline
Wales lidar data & LIDAR terrain and surfaces models Wales & 16-09-2021 & 2m resolution data was used and accessed using EDINA LIDAR Digimap Service\textsuperscript{4}\\ 
\hline
\end{tabular}
\\
\begin{flushleft} 
Note: In regions where lidar data was available as part of both the England and Wales lidar datasets, the data values from England were used.\\
\textsuperscript{1} \url{https://www.hikr.org/region516/ped/?gps=1}\\
\textsuperscript{2} \url{https://www.hikr.org/region518/ped/?gps=1}\\
\textsuperscript{3} \url{http://zverik.openstreetmap.ru/gps/files/extracts/europe/great_britain.tar.xz}\\
\textsuperscript{4} \url{https://digimap.edina.ac.uk}\\
\textsuperscript{5} \url{http://download.geofabrik.de/europe/great-britain.html}\\
\end{flushleft}
\label{}
\end{adjustwidth}
\end{table}

\section*{S2 Supporting Information. Data acquisition and preparation}

\subsection*{Importing GPS tracks}
\label{1Importing GPS tracks}

Approximately 20,000 tracks with GPS data were obtained, the majority of the tracks used here were from OpenStreetMap (OSM) \cite{OpenStreetMap.org2021Tracks}, with a second, smaller set from Hikr.org \cite{Hikr.org2021UnitedReports}. The OSM data consisted of a GPS data dump of all tracks on OpenStreetMap which contain points within the UK, as of April 2013 (the date of the last GPS data dump by OSM). This dataset does not contain any information about the mode of transport being used when the track was recorded, or the device used for the recording. The Hikr dataset contained all of the Hikr reports which were uploaded from July 2009 to July 2021. Although only relatively small quantities of Hikr data are available, all of the tracks used were explicitly tagged as hiking reports, so could be used as the basis of a filter to determine which of the OSM tracks contained walking data (see Data Filtering). 

Each GPS track is made up of track segments, where each track segment is a continuous run of points; a point is a data pair containing the device location and time of recording. The exact nature of tracks and track segments depends on the individual device settings being used. Some devices record points at fixed time intervals, others may only record a new point after travelling a sufficiently large distance away from the previous point. In general, a new track or a new track segment is created whenever the device is switched off or loses signal for a period, although this is not always the case.

All track segments which were fully contained within the area covered by the OS Terrain 5 DTM were imported (see Elevation and Slope). The GPX files were read using a customised version of the \texttt{GPX Segment Importer} plugin for QGIS \cite{SGroe2019QGISImporter}, extended to also read elevation and terrain information (code on Github - \texttt{\href{https://github.com/AndrewWood94/PhDThesis}{AndrewWood94/PhDThesis}}). Although a single GPX file can contain multiple separate tracks, we considered all data within a file to be part of the same track. Tracks within the same file were highly likely to be undertaken by the same individual, and thus have correlated walking speeds. To account for the variance caused by different hike locations and difficulty, each track segment within a file was processed individually.

For each track segment, the list of points was converted to a series of connected linestrings, with the following properties calculated and attached to each one:

\begin{itemize}
    \item Start coordinate
    \item End coordinate
    \item Start time
    \item Duration
    \item Distance
    \item Speed
\end{itemize}

We took a decision to exclude very short tracks (less than 250m in length of 2.5 minutes in duration as well as any tracks where the median speed was greater than 10km/h, as it was felt they might not be a reliable representation of a real walk.

Duplicate track segments (those which contained a section with the same start and end coordinates, the same starting date and time, and the same duration) were tagged, and only the first instance of each was retained. It is believed that these duplicates were a result of either users uploading the same track multiple times, or GPS devices recording multiple versions of the same track with different levels of accuracy or automatic filtering. Finally, a small number of outlier Hikr tracks were checked manually and removed as looking at the metadata or track description revealed that the routes in question were labelled as trailruns, and so should not be considered.

\subsubsection*{Elevation and Slope}
\label{1Elevation and Slope}

The Ordnance Survey Terrain 5 Digital Terrain Map (DTM) \cite{OrdnanceSurvey2020Terrain5} provides the elevation over the whole of Great Britain at 5 m intervals, with an accuracy of greater than 2.5 m RMSE. After reading in each track, the DTM was used to calculate the following:

\begin{itemize}
    \item Elevation
    \item Walking slope
    \item Hill slope
\end{itemize}

Although the GPX files also contained elevation data which could be used to calculate the walking slope, there were a number of reasons to prefer the Ordnance Survey data:

\begin{itemize}
    \item Multiple tracks were found where the elevation was measured to the nearest half metre or metre on the GPS device. Using this value would lead to rounding errors in slope calculations over short distances. The Ordnance Survey values for elevation have a higher level of precision which could alleviate these errors.
    \item Using OS values avoids any concern about potential discrepancies arising from differences in calibration across GPS devices, so we could ensure that the same location on different tracks would have the same elevation.
    \item We were also investigating the impact of hill slope, which could not be calculated from the GPS track data, and using the same data source for both calculations should allow for better evaluation of interactions between the two variables.
\end{itemize}

When first reading each data track, the elevation DTM was sampled to provide the spot height of the start co-ordinate, and the hill slope at this point was calculated using the quadratic surface method \cite{Zevenbergen1987QuantitativeTopography, Dunn1998TheGIS}. As explored in previous work \cite{Jones1998ADEM}, this method produces the most accurate slope estimates, especially given the high resolution of the DTM being used for calculations. The walking slope was calculated using the spot heights of the start and end coordinates of each linestring, and the distance between those points.

\subsubsection*{Roads and Paths}
\label{2Roads and Paths Description}

Road and path data from OpenStreetMap was downloaded from GeoFabrik.de \cite{OpenStreetMap.org2021Data} and contains the data as of August 2021. Although similar data is available from Ordnance Survey, the OSM dataset was preferred due to it providing a more detailed classification than Ordnance Survey, in terms of the road or path type. 

The OSM road and path data is made up of linestrings (single vector lines, as opposed to area features which match the width of the road) and it would be very unlikely for all of the points on a GPS route which follows a road to fall exactly along that line, making it difficult to determine which points should be classified as following the road or path. This problem was then exacerbated by two factors; inaccuracies within the GPS readings, and inaccuracies within the map data itself. This combination of potential errors meant that we needed to decide upon a radius around each point to search for roads or paths.

Here we classified a point as being on-road if a single feature of the OSM road dataset was found within a 50 m radius around the point. This radius was determined after a number of iterations. 

This inclusive distance will lead to an over classification of on-road points. Due to the relative numbers of on-road vs off-road points, we decided it would be preferential to mis-classify points as being on a road rather than the other way around, as this would have a much smaller impact on any resulting models.

By consulting the descriptions of the OSM road-type definitions \cite{OpenStreetMapWikiRoads}, we could separate the on-road points into two categories, paved and unpaved. When doing this, we assumed that the standard road type was paved. Therefore, if a point contained multiple road types, we only considered it to be unpaved if none of the road types detected were paved. The paved road types are the following:

 \begin{multicols}{3}
\begin{itemize}   
    \item Cycleway       
    \item Footway       
    \item Living\textunderscore street  
    \item Motorway       
    \item Motorway\textunderscore link 
    \item Pedestrian     
    \item Primary      
        \columnbreak
    \item Primary\textunderscore link   
    \item Residential    
    \item Secondary    
    \item Secondary\textunderscore link 
    \item Service   
    \item Steps  
    \item Tertiary   
    \columnbreak
    \item Tertiary\textunderscore link  
    \item Trunk        
    \item Trunk\textunderscore link     
    \item Unclassified   
    \item Unknown
    \item[]
    \item[]
\end{itemize}
\end{multicols}

While these are unpaved:
\begin{multicols}{3}
\begin{itemize}
    \item Bridleway   
    \item Path 
    \item Track       
    \columnbreak  
    \item Track\textunderscore grade1   
    \item Track\textunderscore grade2  
    \item Track\textunderscore grade3 
    \columnbreak  
    \item Track\textunderscore grade4   
    \item Track\textunderscore grade5
    \item[]
\end{itemize}
\end{multicols}

\subsubsection*{Obstruction Height}
\label{2Obstruction Height}

For England and Wales, a lidar Digital Terrain Map (DTM) and Digital Surface Map (DSM) were downloaded (see \nameref{S1_Appendix}), both at 2 m resolution, with an accuracy of 15 cm RMSE \cite{LIDARDSMEngland, LIDARDTMEngland, LidarWales}. The DTM gives the ground height above sea level every 2 m, while the DSM provides the surface height (i.e. taking into account buildings or trees etc). For Scotland, there is limited coverage in more rural (i.e. off-road) areas, and most of the available data is at very high resolution (25-50 cm). This is generally greater than the accuracy of the GPS devices used. Given the partial coverage and resolution mis-match in the Scottish datasets, further analysis was limited to tracks in England and Wales.

The start coordinate for each point in a GPS track was sampled in both the DTM and DSM, and the difference between the two values taken as the level of terrain obstruction for the point. For example, a point in a woodland may have a DTM height of 80 m above sea level, and a DSM height of 85 m (the height at the top of the tree canopy), giving us a terrain obstruction value of 5 m.

\subsection*{Break Finding}
\label{1Breakfinding}

For this work, we were interested in calculating the active components of a route (as opposed to breaks or rests), as that is the data on which a walking speed model should be based. Performing analysis on the data without first removing breaks would likely result in inaccurate and lower estimates for the movement speed. Similarly, when developing the filter to find walking routes in the OpenStreetMap dataset we wanted to ensure that we were only considering the active route components.

Initial inspection of the walking tracks showed that a large number contained obvious breaks, while the device was still recording. The simplest method to find breaks, and the first implemented, was to tag all points where there was no movement (i.e. walking speed = 0 km/h). The next step was to tag any individual points which represented over 1 km or 3 minutes of travel as breaks. These points could occur in areas where the device lost signal for a period. Although this should result in the creation of a new track segment, there were a number of tracks where this was clearly not the case. Points with speeds \textgreater10 km/h occurring immediately following a long (\textgreater3 minute) point were also tagged as breaks. Situations like this occurred on a number of devices, likely when a device automatically paused recording for a break. Once significant movement was detected, two points were then added in quick succession, the first at the original location with the time the movement started again, and the second at the new position of the device. This resulted in one high duration point with very little movement, followed by single-second duration movement with high speed, where the device `caught up' with the correct location.

Unfortunately, this did not capture all of the break points due to GPS drift; the distance between the measured position of the GPS device and the true location. This error is always present, but is more obvious when stationary, as a large number of points are recorded in the same general area, forming clusters. Examples of these clusters can be seen in Fig \ref{Fig2}.

Although these locations are easy to identify when the route is visualised, we needed an automatic filter in order to remove them from our analysis. The movement speeds calculated from these drift points can vary greatly, depending on the sampling rate of the device and the amount of GPS drift. A drift measurement of 6 m in a very short time (1-2 seconds) would mean a very high speed (10-20 km/h) is found, but a series of very low speeds in a row could be due to a break with a small amount of drift, or it could indicate a particularly difficult (and therefore slow) section of the route. 

A number of previous studies explored methods to classify GPS tracks into activity types \cite{Zhou2017AData, Schuessler2009ProcessingInformation, Biljecki2010AutomaticModes, Tsui2006EnhancedSystems, Wan2016ClassifyingApproach, DeVries2012FilteringClassification, Alvares2007AInformation, Palma2008ATrajectories}, however none could be directly applied to this problem. This is because they were usually trying to identify different modes of transport, which can be clearly distinguished by different travel speeds. However, as mentioned above, the speed measurements caused by GPS drift can easily be in the same range as an expected walking speed. Methods which look for clusters similar to those being investigated here were also not useable as they generally require a known sampling rate for the GPS device \cite{DeVries2012FilteringClassification}, or involve checking clusters against a pre-existing database of features \cite{Alvares2007AInformation, Palma2008ATrajectories}. Our data consisted of tracks created using a wide range of devices and settings, and over a very large area, so it was not possible to either assume a fixed sampling rate, or to pre-select features where breaks were likely. Instead, various ideas from a number of works were combined and adapted to find clusters of points, which were then checked to see if they should be identified as a break.

Firstly, based on \cite{Palma2008ATrajectories}, a modified version of DBSCAN \cite{Ester1996ANoise} was used to identify clusters. Unlike in the standard algorithm, each point only looked for neighbours whose timestamp was within 10 minutes of the point being checked. This prevented clusters being found on routes which doubled back on themselves or contained loops. Secondly, as sampling rates were not consistent across devices and the tracks covered a variety of terrains, we could not assume a fixed radius to find neighbouring points. Instead the median travel distance for the particular track segment being investigated was used (r\textsubscript{median}).\\

\begin{quote}
    \textit{DEFINITION 1. Neighbourhood of a point: Let $\{p_0, p_1,...,p_n\}$ be points on a GPS track segment, with timestamps $\{t_0, t_1,...,t_n\}$, a median distance $r_{median}$ between consecutive points and a median point speed $s_{median}$. The neighbourhood $N_k$ of a point, $p_k$, is the set of points $p_i$ such that: $$dist(p_i, p_k) < r_{median} \text{ and } |t_i-t_k| < 600 s$$}\\
\end{quote}

Using these conditions, all points along each track segment were tested to find point clusters.\\
 
\begin{quote}
    \textit{DEFINITION 2. Point cluster: A point cluster, C, is formed from a point neighbourhood, $N_k$, if any of the following hold:
    \begin{enumerate}
        \item At least 5 non-consecutive points are found in $N_k$
        \item At least 10 consecutive points are found in $N_k$
        \item A point within $N_k$ has a `high speed'; a speed greater than $2*s_{median}$
        \item A point $p_k$ is immediately preceded by a `high speed' point
        \item A point within $N_k$ has a `very low speed'; a speed less than 0.01 km/h\\
    \end{enumerate}
    }
\end{quote}

\begin{figure}[!h]
    \begin{adjustwidth}{-1.75in}{0in} 
\includegraphics{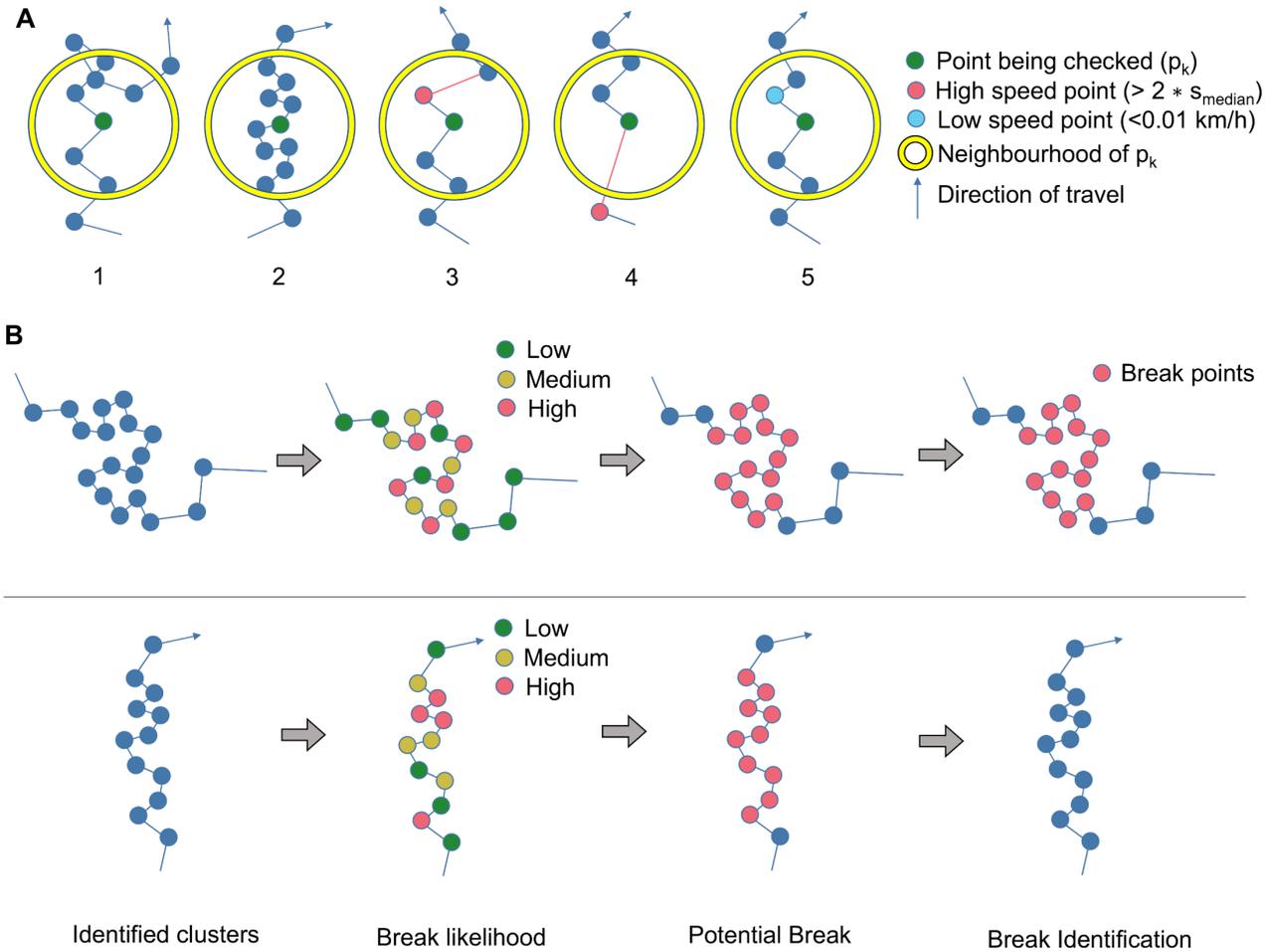}
    \captionsetup{width=1\linewidth}
    \caption[width=\textwidth]{{\bf Demonstration of how breaks are identified from point clusters.} (A) The five scenarios in which a point cluster is identified around a point. (B) Points in a cluster are checked to find their break likelihood, and see if a potential break can be identified. Potential breaks are then checked to see if a break can be formed.}
    \label{Fig8}
    \end{adjustwidth}
\end{figure}

Examples of each of these conditions are shown in Fig \ref{Fig8}A. The first two conditions worked together to prevent finding clusters in unusually slow sections of a route, such as a steep hill climb, but allowed for areas where GPS drift was small and an entire cluster was contained within $N_k$. The third and fourth conditions found `high speed' points which occurred when a large amount of GPS drift was measured, and could spread a cluster out over a wider area. Without accounting for these points separately, the algorithm would often end up registering a single break as multiple short breaks separated by high speed movements. The fifth condition was necessary for situations where the device was set to only record new points once a minimum distance has been travelled from the previous location. Note that in the third and fifth cases in definition two, the point immediately following the high- or low-speed point was added to the cluster, even if it was not included in $N_k$. Similarly, in the fourth case, the cluster included the preceding high-speed point regardless of whether it was in $N_k$.

Once a point cluster was identified, each point within it was tested and any further clusters found were added to the original. This continued recursively until every point within the cluster had been checked and no new clusters were found. This allowed us to identify clusters (and breaks) which lasted longer than the 10 minute Neighbourhood threshold, as each point added to a cluster allows the search window to be extended beyond the initial limit. 

Following this, further steps were taken to try and ensure that only legitimate breaks were classified, rather than slow sections of movement. Firstly, each point within the cluster was tested for `break likelihood' using a simple classification methodology based on the approach used by Wan and Lin \cite{Wan2016ClassifyingApproach}, which uses two variables; the point speed, $s_i$, and the point angle, $\alpha_i$.\\ 

\begin{quote}
    \textit{DEFINITION 3. Point speed: The speed, $s_i$, at a point, $p_i$ is categorised as follows:
    \begin{itemize}
        \item[--] Low: $s_i < s_{median}/2$
        \item[--] Medium: $s_{median}/2 <s_i < 10 m/s$
        \item[--] High: $s_i > 10 m/s$\\
    \end{itemize}
    }
\end{quote}

\begin{quote}
    \textit{DEFINITION 4. Point angle: The point angle, $\alpha_i$, for a point, $p_i$, is the angle created at $p_i$ by the lines connecting it to $p_{i-1}$ and $p_{i+1}$. It is then categorised as follows:
    \begin{itemize}
        \item[--] Narrow: $\alpha_i < 90^{\circ}$
        \item[--] Wide: $\alpha_i >= 90^{\circ}$\\
    \end{itemize}
    }  
\end{quote}

As discussed in \cite{Wan2016ClassifyingApproach}, normal walking is unlikely to result in point angles below 90 degrees, unlike a stationary object subject to GPS drift which often produces narrower point angles. The break likelihood for each point in a cluster was found using the classifications in Table \ref{tab:1Break Likelihoods}. 

\begin{table}[!h]
\centering
\caption{Break likelihood classifications based on point speeds and angles}
\begin{tabular}{|c|c+c|c|c|} 
\hline
\multicolumn{2}{|c+}{} & \multicolumn{3}{c|}{\textbf{Point Speed}}  \\
\cline{3-5}
\multicolumn{2}{|c+}{} & Low & Medium & High\\ 
\thickhline
\multirow{2}{*}{\textbf{Point Angle}} & Narrow & High & Medium & High\\ 
\cline{2-5}
 & Wide & Medium & Low & Medium \\
\hline
\end{tabular}
\label{tab:1Break Likelihoods}
\end{table}

Once the break likelihoods were identified, the cluster was checked to find a potential break. Performing this step helped to limit the sizes of breaks, as walking points immediately preceding or following a break were often caught in the cluster.\\

\begin{quote}
    \textit{DEFINITION 5. Potential Break: A potential break, $B^\ast$, is created by ordering the points in a cluster and identifying the first and last points with a break likelihood of medium or high. All points in the cluster between these points (including the points themselves) form the potential break.}\\
\end{quote}

As a break by definition implies no movement, any GPS drift in a given direction should be cancelled out by subsequent GPS drift in the opposite direction. 
For this reason, the bearing of all points within the potential break was found and assigned a quadrant, and a break was only formed if there was apparent motion in opposite quadrants.

\begin{equation*}
\begin{aligned}
   Q_i = \ &True \ iff \\ 
   &\exists \ p \in B^\ast \ | \ (90*(i-1))^{\circ} <p_{bearing} < (90*i)^{\circ}, \\ 
   &i=(1,2,3,4)
\end{aligned}
\end{equation*}

\begin{quote}
    \textit{DEFINITION 6. Break: A break, $B$, is created from a potential break, $B^\ast$, if both of the following hold:
    \begin{itemize}
        \item $Q_1 = True \ and \ Q_3 = True, \ or \ Q_3 = True \ and \ Q_4 = True$.
        \item Less than half the points in $B^\ast$ have a low break likelihood.\\
    \end{itemize}
    }
\end{quote}

Fig \ref{Fig8}B provides examples of two point clusters and the process to identify breaks. Note that not every point in each cluster becomes part of the potential break; low break likelihood points at the ends of a cluster are excluded. After processing the second cluster, a break is not defined because the cluster does meet the requirement of apparent travel in opposite quadrants. While there is a substantial amount of east-to-west deviation in the track, it consistently heads north. Regions such as this identified in GPS tracks are more likely to indicate a period of challenging (and therefore slow) movement. 

A substantial ground truth dataset of routes where the breaks are tagged does not exist, so we were unable to measure the accuracy of this break-finding algorithm numerically. However, a qualitative inspection of a small number of tracks suggested that it classified breaks well, with over-classification in some circumstances. This was preferable to under-classification for two reasons;  firstly, as discussed below, only the longer breaks were actually removed from calculations so small breaks found as a result of over-classification have no impact. Secondly, we were performing analysis on very large datasets, so we expect the analysis to be robust to the incorrect removal of a small number of data points.

\subsection*{Data Filtering}
\label{1DataFiltering}

Before processing our data, we wanted to use information about walking speeds found in the Hikr data to identify and remove non-walking tracks from the OSM dataset. When doing this, we only wanted to compare the active component of each of the tracks. However, not all break points should necessarily be considered `inactive'. The majority of the breaks seen in the dataset were under 30 seconds in length, and these were defined as `micro-breaks'. Micro-breaks were felt to be a constituent part of the walk which most people would have to do, such as pausing to catch your breath, and were not excluded from further analysis. The exception to this was if the micro-break contained a single point with over 1 km of movement, or points with speeds of over 10 km/h. Speeds over 10 km/h in a region which had been identified as a break were very likely to be errors caused by GPS drift, and as such these points would not be useful in representing a typical walking speed. It was felt that 30 seconds is a reasonable length such that breaks of this length or longer would be a conscious choice by the walker to stop, rather than being necessary for the route. After excluding long breaks, it was also decided to remove any breaks which occurred at the start or end of a track segment.

The remaining data were merged together into continuous sections at least 50 metres in length, to remove some of the variability caused by the GPS devices and elevation data resolution. Devices generally measure the time to the nearest second, so rounding errors over short distances could have a large impact on the estimated speeds, especially in combination with small inaccuracies in the location data. By merging the data together into longer sections the impact of these errors was greatly reduced, as any error made up a smaller percentage of the merged section. A similar inaccuracy existed with walking slope values. As our elevation DTM had a resolution of 5 m, a distance of under 5 m between two consecutive points could result in a walking slope of 0 degrees, regardless of how steep the terrain is in reality. By merging the points together, we smoothed out the steps in the data and produced a single datapoint with a slope value closer to the true value, which will help produce a model with more accurate walking speed predictions. 

When merging data, the distance was taken as the cumulative distance across all points, and not the direct start point to end point distance. The merged point was classified as on a paved road if at least one of the constituent points was classified as being on a paved road. (Similarly a point was only classified as being off-road if none of the constituent points were on-road). The hill slope, walking slope and average obstruction height were calculated as the weighted average of the value at each datapoint, weighted by the duration of each point.

Once the data were merged into 50 m sections, any section with a speed above 10 km/h which was at the start or end of a segment, or next to a break point was considered to be part of the break. This was repeated recursively until no more high speed points were found next to breaks.

Although all of the Hikr data was tagged as a walk or hike, there were a small number of individual track segments with high average speeds. Upon further inspection it was clear that these were segments, within a larger walking track, where other modes of transport were used and should be filtered out, therefore segments with an average speed greater than 10km/h were removed.

For each remaining track segment the median, upper quartile and maximum speed were calculated, and statistics from these were found to use in filtering the OpenStreetMap data to remove non-walking tracks:

\begin{align}
    \begin{split}
        &\text{The upper quartile of the maximum Hikr speeds} \\ &\text{(approx. 6.0 km/h)}\label{Q1Hikrmax} 
    \end{split}\\
    \begin{split}
        &\text{The median of the median Hikr speeds} \\ 
        &\text{(approx. 3.2 km/h)}\label{medHikrmed}
    \end{split}\\
    \begin{split}
        &\text{The upper whisker of the maximum Hikr speeds} \\ &\text{(approx. 8.1 km/h)}\label{topHikrmax}
    \end{split}\\    
    \begin{split}
        &\text{The minimum of the upper quartile Hikr speed}  \\ &\text{(approx. 2.3 km/h) }\label{minHikrQ1}
    \end{split}
\end{align}

The OpenStreetMap data were also merged into intervals of at least 50 m, while ignoring short micro-breaks. Similarly to the Hikr tracks, this dataset contained a number of tracks where there were clearly multiple movement methods, often when the user was driving or cycling to a hike location. Unlike in the Hikr data however, a change in transport method was often not accompanied by the start of a new track segment. Instead, individual route segments appeared to contain a variety of transport methods, often separated by a number of very extreme points (likely where the device lost signal for a period). As we did not want to remove these segments entirely if they contained valid walking data, we used the extreme points as markers to break the segment down into smaller sections. The `key points' were defined as the start and end points of the segment, as well as any point with a distance greater than 500 m, a duration greater than 3 minutes, or an apparent speed over 100 km/h (this is typically caused by a device refresh). After identifying the key points, the following conditions were applied:

\begin{itemize}
    \item If only a single data point existed between a pair of key points it was ignored
    \item If the median speed between a pair of key points was greater than (\ref{Q1Hikrmax}), then all points in the range were ignored
\end{itemize}

Following this, all segments were checked and the steps outlined below were carried out to remove unwanted data. These were repeated until no further data was removed.

\begin{enumerate}
    \item If the segment contained less than 2.5 minutes or 250 m of useable data it was removed
    \item Segments were removed if any of the following were true:
        \begin{itemize}
            \item The median speed was greater than (\ref{Q1Hikrmax})
            \item The minimum speed was greater than (\ref{medHikrmed})
            \item The upper quartile speed was greater than (\ref{topHikrmax})
            \item The upper whisker speed was less than (\ref{minHikrQ1})
        \end{itemize}
    \item All points with a speed above 10 km/h which were at the start or end of a segment, or next to a break point were considered part of the break.
\end{enumerate}

For the remaining segments, the Hikr and OpenStreetMap data were then combined together into a single dataset. Visual inspection of the dataset revealed that the filters were good, but a small number of outliers remained. Therefore, the fastest and slowest 0.5\% of the merged datapoints were removed (classified as breaks for further processing) and were not included in any modelling.

Following this a decision was made to remove data from tracks found in Scotland. Lidar data covering the walking tracks was necessary to model the terrain obstruction, and was not sufficiently available in Scotland at the time of the study. Furthermore, analysis showed that that walking speeds in Scotland were at the extreme end of what is seen throughout the rest of the UK (see \nameref{S4_Appendix}). Including this data without also including a corresponding extreme dataset where lidar data is available may result in incorrect modelling. All OSM track segments which took place within Scotland were excluded from further processing. Similarly Hikr tracks which were tagged as taking place in Scotland, and which fully took place in Scotland were excluded. Note that we defined tracks in Scotland as any track segments which were fully contained within the following OS grid squares: 

            HP
         HT HU
   HW HX HY HZ
NA NB NC ND   
NF NG NH NJ NK
NL NM NN NO
   NR NS NT NU
   NW NX NY 
   
Within the NY tile, the following tiles were excluded:
 
09 19 29 39 49 59 69
08 18 28 38 48 58 
07 17 27 37 47 
06 16 26 36 

A small number of Hikr tracks which took place in the Scottish islands were not removed as these were tagged separately on Hikr.org.

This left us with a final dataset of almost 88,000 km and over 7,600 tracks across the UK, with a mean speed of 4.64 km/h. All of the data remaining was assumed to consist solely of hiking or walking tracks, although there are likely to be a number of areas where this was not the case. There is not a large difference in speed profile when walking or cycling up a steep incline, so further data filtering would require in-depth analysis of each individual track. There was no objective way to remove invalid points without potentially removing valid data as well. However, the volume of data used for modelling should alleviate errors arising from any non-hiking track segments remaining.

\section*{S3 Supporting Information. Exploratory data modelling study.}
\label{1Modelling}

Data from Scotland was used for exploratory testing of different modelling approaches. When doing this, an earlier iteration of the break finding and data filtering process was used. Data were processed as described in \nameref{S2_Appendix}, with the following exceptions:

\begin{itemize}
    \item GPS track segments had to be fully contained in the reduced `Scotland' OS grid tiles described in \nameref{S2_Appendix}.
    \item Track segments where the median speed was greater than 10 km/h were not automatically removed prior to break identification.
    \item  Individual points representing 10 minutes of travel were tagged as breaks (3 minutes was used in later versions).
    \item High speed (\textgreater10 km/h) points occurring immediately following a long (\textgreater3 minute) point were not automatically tagged as breaks.
    \item When merging data into 50 m sections, sections under 50 m in length immediately preceding a break or the end of the segment were also tagged as breaks, rather than combined with the previous section.
    \item After merging the data into 50 m sections, any section with a speed above 10 km/h which was at the start or end of a segment, or next to a break point were only considered to be part of the break in the OSM data, not the Hikr data.
    \item A minimum distance of 250 m of travel was included between breaks. Sections with a distance below this were tagged as a break.
    \item When looking for `key points' to filter out non-walking track sections, the duration required to be tagged as a `key point' was 10 minutes (3 minutes was used in later versions).
    \item Duplicate track segments were not identified or removed.
    \item The fastest and slowest 0.5\% of the merged datapoints were not removed as outliers.
\end{itemize}

Once the data were filtered and processed, we were able to use them to test models for the walking speed. Two different approaches were explored in order to model the data: a generalised linear model (GLM) and a generalised additive model (GAM).

We know that predictions for walking speeds must be non-negative, and two different setups were explored to achieve this: a Gaussian distribution with log link function and a Gamma distribution with inverse link function. The GAM approach was also deployed with both thin plate spline or cubic regression basis functions.
Investigations into different models showed that there was no improvement to model fit beyond cubic terms in a GLM, or 7 knots in each GAM smoothing term, so more complex models than this were not considered for selection.

Both model types were created in R version 3.6.1:

\begin{equation*}
    \begin{aligned}
    &\texttt{glm}(v \sim a\phi + b\phi^2 + c\phi^3 + d\theta + e\theta^2 + f\theta^3, \texttt{distribution}) \\
    &\texttt{gam}(v \sim s(\phi,k,\beta) + s(\theta,k,\beta), \texttt{distribution})\\      
    \end{aligned}
\end{equation*}

where
\begin{quote}
$v = \text{walking speed}$\\
$\phi = \text{hill slope angle (degrees)}$\\
$\theta = \text{walking slope angle (degrees)}$\\
k = knots used in spline (up to 7)\\
$\beta$ = basis function\\

\end{quote}

Initially, 10-fold cross-validation was used to compare the model parameters, looking at R-squared values, root-mean-squared error (RMSE) and mean absolute error. All models produced very similar results, with no change in the RMSE to 2 decimal places, although there was a general trend of marginal improvements as the model complexity increased. As no best model could be chosen based on the cross-validation, each was checked in more detail. Firstly, the hill slope component was isolated by investigating the speed when the walking slope was zero (i.e when traversing across a slope). Intuitively, and from experience, this should be a decreasing function; as the slope gets steeper it is harder to traverse, so the walking speed will decrease. Models which failed to predict this were removed under the assumption that the data were overfitted. Following this, the walking slope component was investigated, specifically looking at the walking speed when travelling directly up- or down-hill. By inspection of the data, existing functions, and intuition, this should be modelled as a roughly bell-shaped function with the peak at, or close to, 0 degrees. Any models which predicted an increase in speed as walking slope steepness increased (from a minimum magnitude of 10 degrees) were removed. Secondly, we know from existing work that there exists a critical gradient at a walking slope of around 15 -- 21 degrees, at which it becomes more efficient to zig-zag up a steep hill rather than going directly uphill. Models which failed to predict the critical gradient occurring below 21 degrees when travelling uphill were also removed.

\begin{figure}[!h]
    \includegraphics{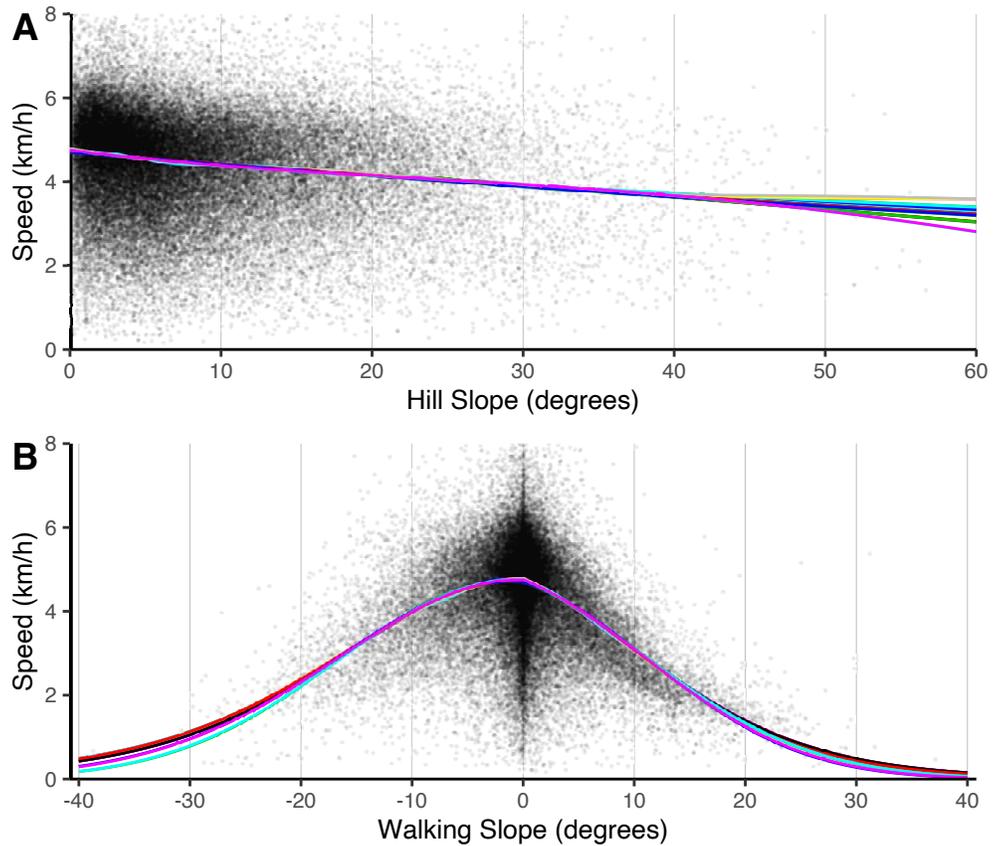}
    \caption{{\bf Walking speed predictions from 21 possible models (coloured individually) generated from the Scotland GPS dataset.} (A) Walking speed predictions for traversing across hills of varying slope, overlaid on GPS data where walking slope is below 5 degrees. (B) Walking speed predictions for travelling directly up or down hills of varying slope, overlaid on GPS data where walking slope is within 5 degrees of hill slope.}
    \label{Fig9}   
\end{figure}

This resulted in 21 model configurations remaining, although it is clear from Fig \ref{Fig9} that the speed predictions are very similar in most circumstances. Fig \ref{Fig9}A shows that all of the remaining models predict very similar speeds when traversing a slope of up to 40 degrees, after which there is more deviation in predictions. Similarly, when travelling in the slope direction (Fig \ref{Fig9}B), all of the models are broadly similar on slopes up to approximately $\pm15$ degrees. More than 96\% of the data is contained within this area, and the relative lack of data outside this region explains the divergent speed predictions. As all of the models provided both very similar R-squared values and very similar predictions over the vast majority of the dataset, we used the following points to make our final selection:
\begin{itemize}
    \item It is easier to apply GLMs than GAMs to future work, as a simple formula to predict the walking speed can be produced for application elsewhere, without needing to recreate the model from the original data.
    \item In general, simpler models are easier to interpret, and we had no clear evidence that a more complex model would perform better.
\end{itemize}

\section*{S4 Supporting Information. Exploring the differences between Scotland and the rest of the UK.}
\label{Scotland VS UK Differences}

\begin{figure*}[!h]
    \includegraphics{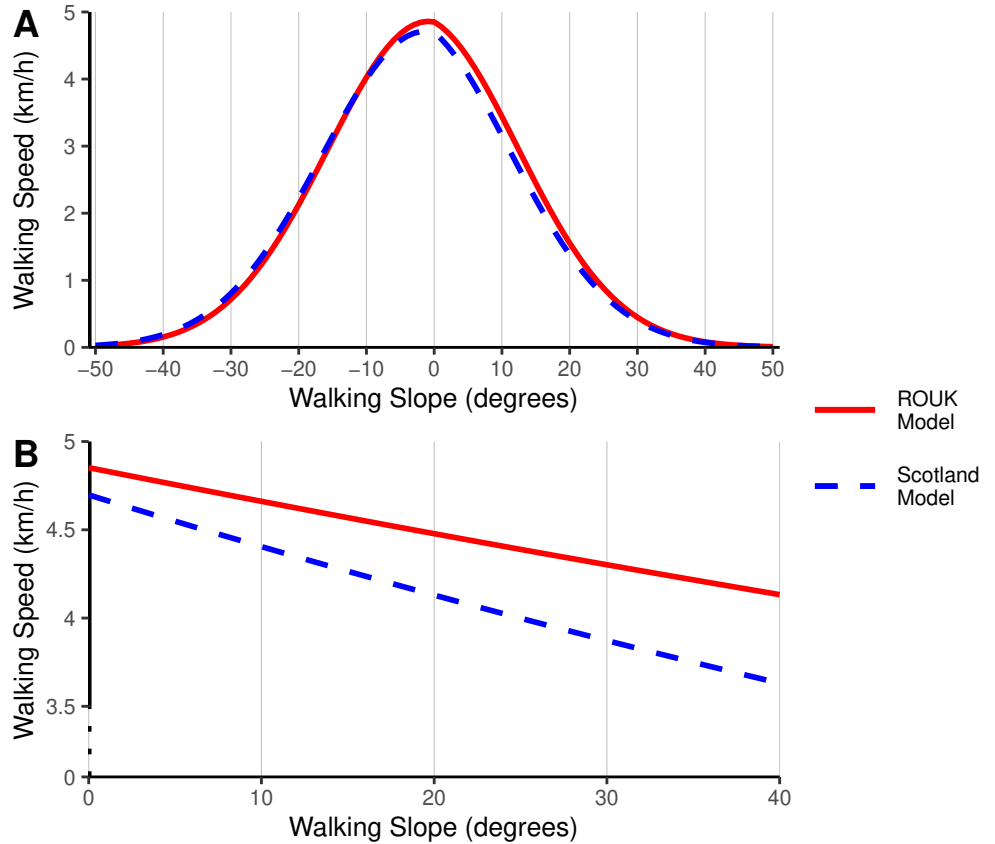}
    \caption{Comparison of walking speed models produced using data from Scotland and the rest of the UK. Walking speed predictions when: (A) travelling directly up or down hills of varying slope, (B) traversing across hills of varying slope.}
    \label{Fig10}
\end{figure*}

When applying the GLM formulation separately to datasets covering Scotland and the rest of the UK (ROUK) (Fig \ref{Fig10}), we see faster predicted walking speeds in the ROUK model than in the Scotland model when traversing a slope, or when walking uphill. Before building a model on a combined dataset, we wanted to check whether the Scotland dataset was a reasonable subset of the ROUK data.

\begin{figure*}[!h]
    \includegraphics{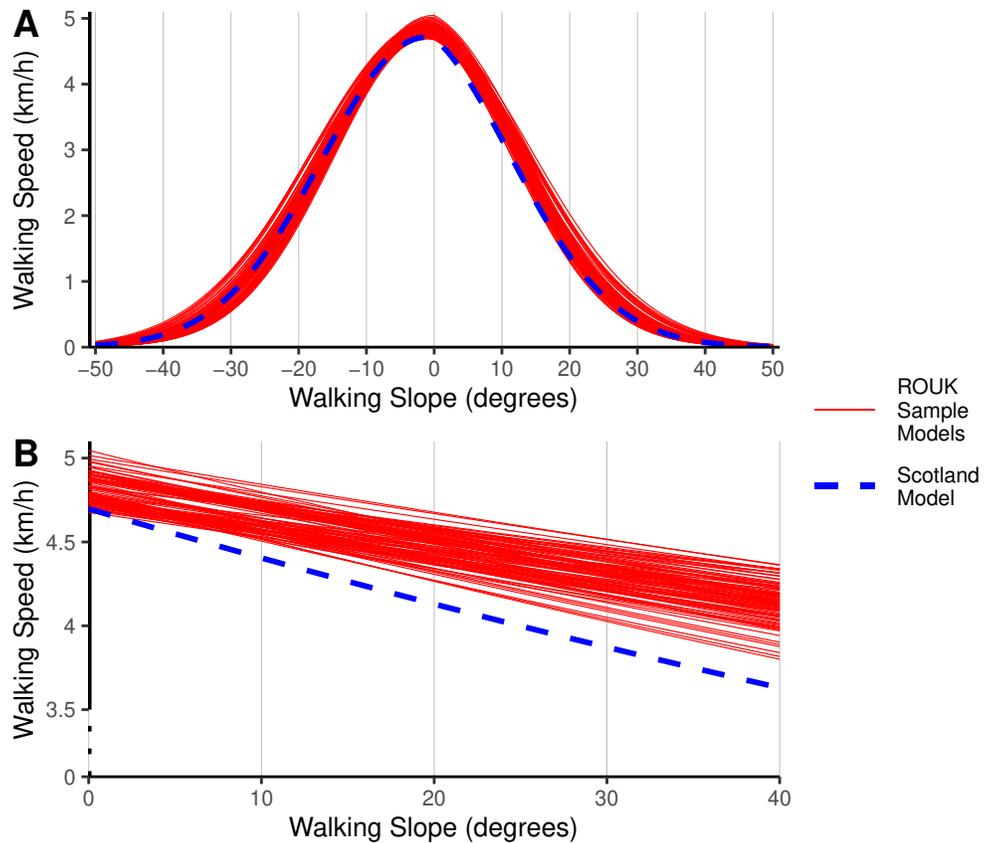}
    \caption{Comparison of walking speed models produced using data from Scotland against 100 sampled datasets from the rest of the UK. Walking speed predictions when: (A) travelling directly up or down hills of varying slope, (B) traversing across hills of varying slope.}
    \label{Fig11}
\end{figure*}

To take into account the fact that the ROUK datset was much larger than the Scotland dataset (7636 tracks vs 648), we took 100 samples of 650 tracks from the ROUK dataset (to form sample sets of comparable size to the Scotland data) and modelled the walking speed for each one. The results are visualised in Fig \ref{Fig11}.

\begin{figure*}[!h]
    \begin{adjustwidth}{-1.75in}{0in} 
    \includegraphics{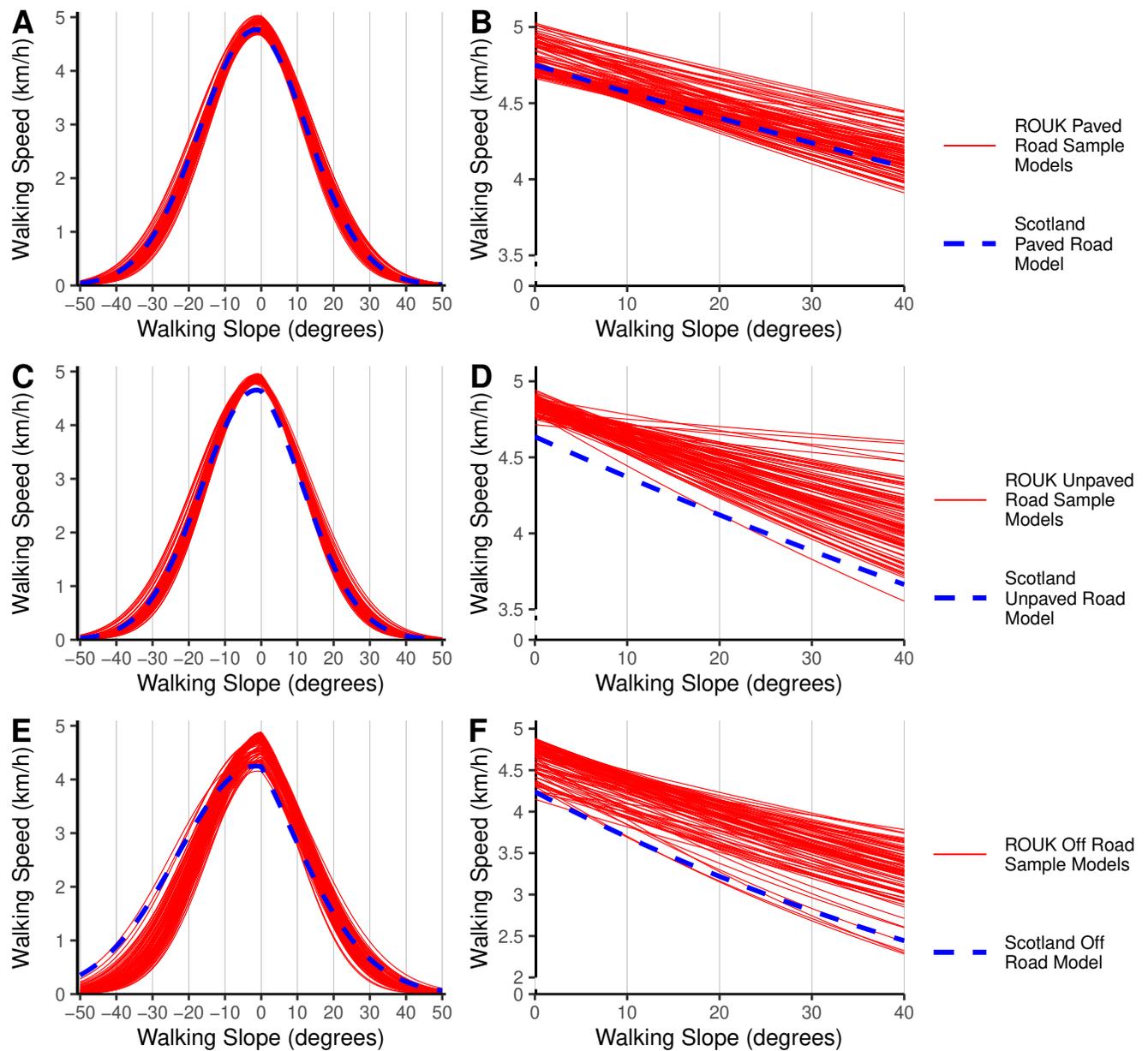}
    \captionsetup{width=1\linewidth}
    \caption[width=\textwidth]{Comparison of walking speed models produced using data from Scotland against 100 sampled datasets from the rest of the UK. Walking speed predictions when: (A) travelling on paved roads directly up or down hills of varying slope, (B) traversing on paved roads across hills of varying slope, (C) travelling on unpaved roads directly up or down hills of varying slope, (D) traversing on unpaved roads across hills of varying slope, (E) for travelling off-road directly up or down hills of varying slope, (F) traversing off-road across hills of varying slope.}
    \label{Fig12}
    \end{adjustwidth}
\end{figure*}

When we look at traversing a hill, it is clear that the two datasets are distinct, as the model for Scotland is outside the range of results seen in the ROUK sample models. Further investigations showed that there were differences between the proportions of paved road, unpaved road and off-road data within each set. To take this into account, we modelled these individually, once again sampling the ROUK dataset 100 times to form sample sets of comparable size to the Scotland dataset (sample sizes of 600 for paved roads and 450 for unpaved and 200 tracks for off-road), and the resulting models can be seen in Fig \ref{Fig12}.

We can clearly see now that our model for paved roads in the Scotland data is comfortably within the range of samples of the ROUK data. It is reasonable to suggest, therefore, that there is no difference in walking on a paved road in Scotland compared to the rest of the UK. However, our unpaved road model and off-raod models for Scotland lie at the extreme edge, or outside of the range of sample models taken from the ROUK unpaved data.

Before finally determining that the two datasets were distinct, we wanted to see if we could find another variable which would account for the lower walking speeds seen in Scotland on both unpaved roads and when off-road. We explored whether this elevation could be responsible for this difference, as higher elevations are likely to have higher exposure, and be more affected by inclement weather, leading to slower walking speeds.

\begin{figure*}[!h]
    \begin{adjustwidth}{-1.25in}{0in} 
    \includegraphics{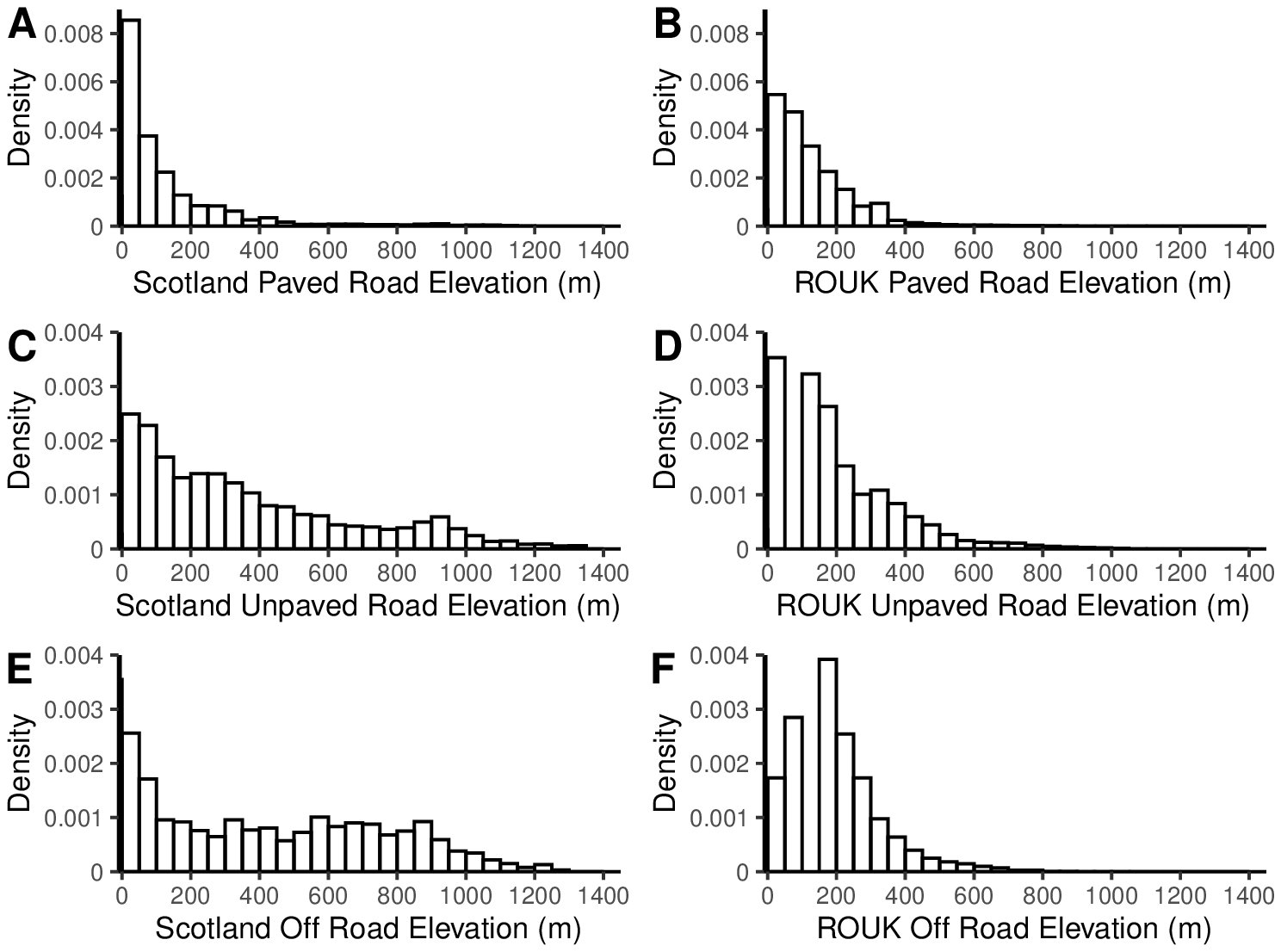}
    \captionsetup{width=1\linewidth}
    \caption{Comparing elevations of track sections between Scotland and the rest of the UK. (A) Scotland paved roads. (B) ROUK paved roads. (C) Scotland unpaved roads. (D) ROUK unpaved roads. (E) Scotland off-road. (F) ROUK off-road.}
    \label{Fig13}   
    \end{adjustwidth}
\end{figure*}

There was a much greater proportion of data at high elevation (\textgreater500 m) in the Scotland dataset than the ROUK dataset on both unpaved roads and when off-road, while a silimar difference was not seen on paved roads (where our models were equal) - Fig \ref{Fig13}.

For this reason we included elevation as a model variable, both as a continuous variable or as a factor variable classifying all points as either high elevation or low elevation (where high elevation consisted of all data \textgreater500 m). However, in both cases we found this to not be a significant factor in the model. Based on the sample data taken, we suggest that the model formulated using the Scotland data is an extreme sample of the ROUK data, where a greater-than-average portion of the data has been sampled from high elevation regions. However, the high elevations themselves are not the cause of the difference between the model coefficients.

\section*{S5 Supporting Information. Exploring the impact of terrain obstruction.}

To understand how terrain obstruction might impact the walking speed (and thus how it should be incorporated into a model), we conducted an initial exploration into the data. Before doing this, however, we wanted to check that there was not a systematic difference between the walking speeds in regions where we had lidar data, and regions where we did not. If the two regions were not found to be different, then any findings about the effects of terrain obstruction in regions where we had lidar data could also be applied to areas where we didn't have the data. 

\begin{figure}[!h]
    \includegraphics{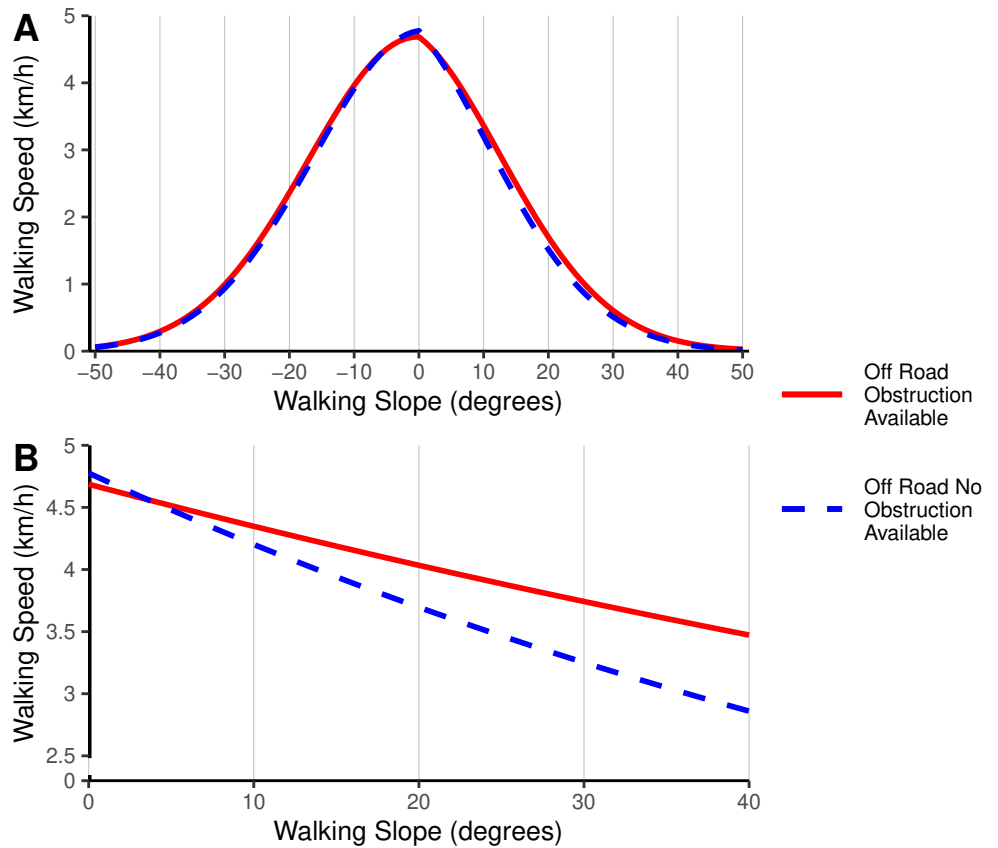}
    \caption{{\bf Comparison of off-road walking speed models where obstruction data is, or is not, available.} Walking speed predictions when: (A) travelling directly up or down hills of varying slope, (B) traversing across hills of varying slope.}
    \label{Fig14} 
\end{figure}

When modelling the data for the separate datasets (`obstruction available' vs `no obstruction available'), we see that the models are very similar when ascending or descending a slope (Fig \ref{Fig14}A). This was not the case when traversing the slope however, as the `no obstruction available' model predicts that hill slope has a greater impact on reducing walking speed than the `obstruction available' data model (Fig \ref{Fig14}B).

\begin{figure}[!h]
    \includegraphics{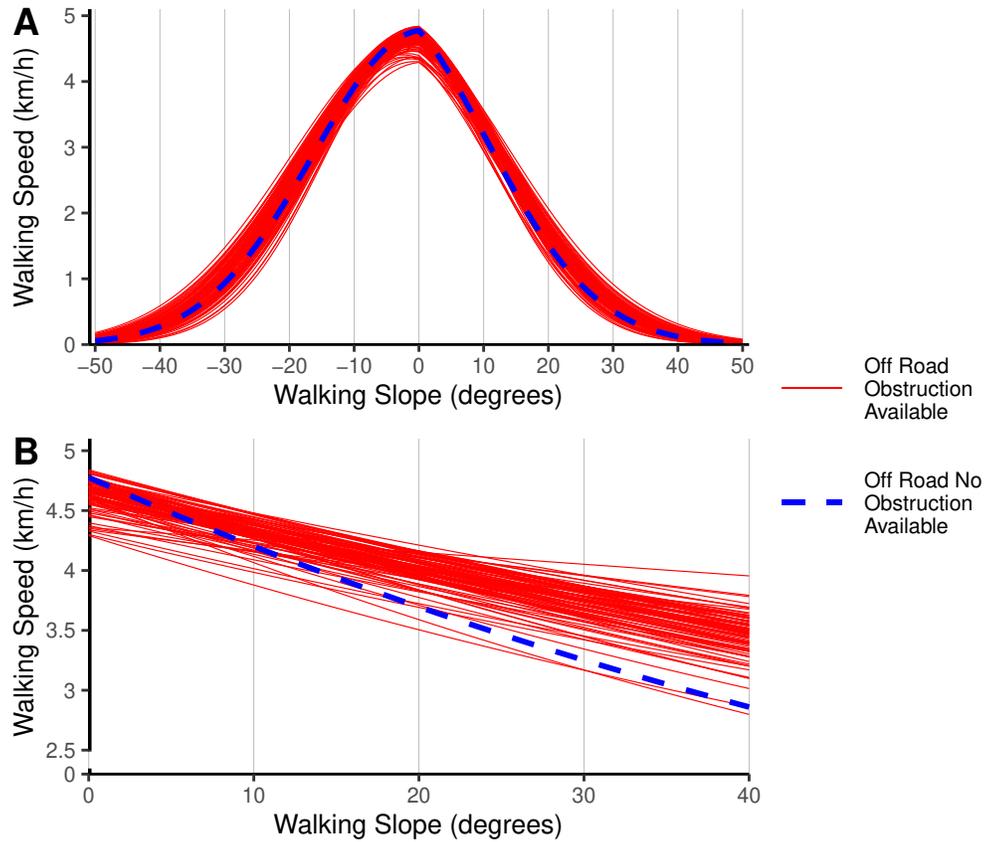}
    \caption{{\bf Comparison of off-road walking speed models produced using a dataset where obstruction data isn't available against 100 sampled datasets where obstruction data is available.} Walking speed predictions when: (A) travelling directly up or down hills of varying slope, (B) traversing across hills of varying slope.}
    \label{Fig15}   
\end{figure}

We sampled our larger (`obstruction available') dataset, so that we had a similar number of tracks as in our smaller dataset, and compared models made from more equal volumes of data. When doing this (Fig \ref{Fig15}), we found that the `no obstruction available' model is within the range of sample models for traversing the slope, albeit at an extreme end. This is likely due to the low volume of data which we had at high hill-slopes. (Only 50 km of data had a hill slope greater than 15 degrees with no lidar data available, and only 130 km with lidar data available). Going forward, we assumed that the regions where we had lidar data were representative of all off-road regions, and so any findings could be applied to both areas.

\begin{figure}[!h]
    \includegraphics{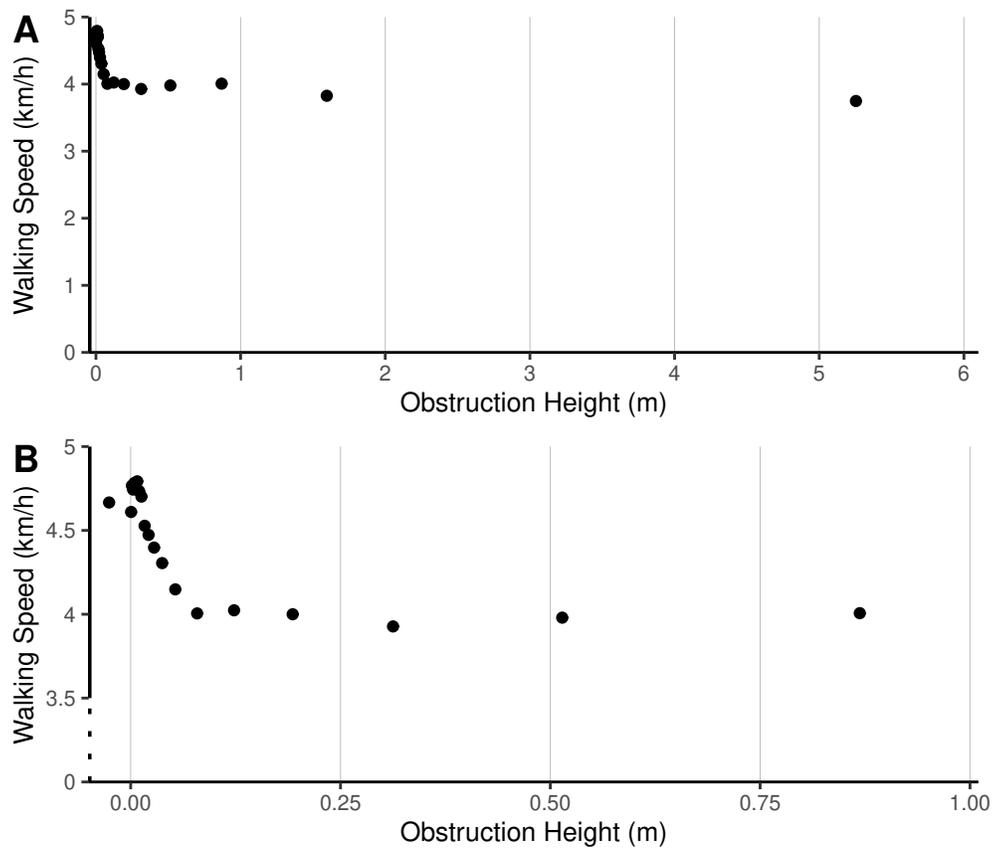}
    \caption{{\bf Binned average walking speeds across different levels of obstruction.} (A) Full range. (B) Zoomed range. Each bin contains 1/25th of the datapoints.}
    \label{Fig16}    
\end{figure}

To explore the effects of terrain obstruction, we first looked at the range of speeds across the different obstruction values. The data was split into 25 quantiles, and the average walking speed for each was calculated. The results are shown in Fig \ref{Fig16}A. This shows us two things; firstly the vast majority of our data had very little, or no obstruction (as most of the quantile points occur below 0.5 m of obstruction). Secondly we can see that there is a very steep drop off in walking speed initially, and it then remains relatively constant across obstruction levels. Our initial assumption was that walking would be relatively easy with no, or very little obstruction, and then much slower at obstruction values of approximately 0.5 m - 4 m when it would involve walking through thick vegetation, before getting slightly faster again at higher obstruction values (as you would be walking through a forest and could walk between the trees below the canopy). The data shows this not to be the case, although this may be a result of our data only showing us regions where walking was possible. Due to the crowdsourced nature of our GPS tracks, we had no data showing us the walking speed when in 2 m of thick vegetation, as it is very unlikely that people would have chosen to walk there.

Fig \ref{Fig16}B shows a close-up of the steep speed drop off, and we can see that the average speed dropped from approximately 4.8 km/h when there was no obstruction down to about 4 km/h once there was more than 10 cm of obstruction. We used this information to classify all points into heavy obstruction (\textgreater10 cm) or light obstruction ($<=10$ cm). Although the figure suggests a gradual decrease in walking speed between 0 and 10 cm of obstruction, we chose not to model this. Vegetation length is highly variable throughout the year, and it is more practical to classify regions as light or heavy obstruction when discussing walking speeds.

\section*{S6 Supporting Information. Comparison of walking speed changes while crossing a simulated off-road terrain region.}

\begin{figure}[!h]
    \begin{adjustwidth}{-1.5in}{0in} 
    \centering
    \includegraphics[width=0.86\linewidth]{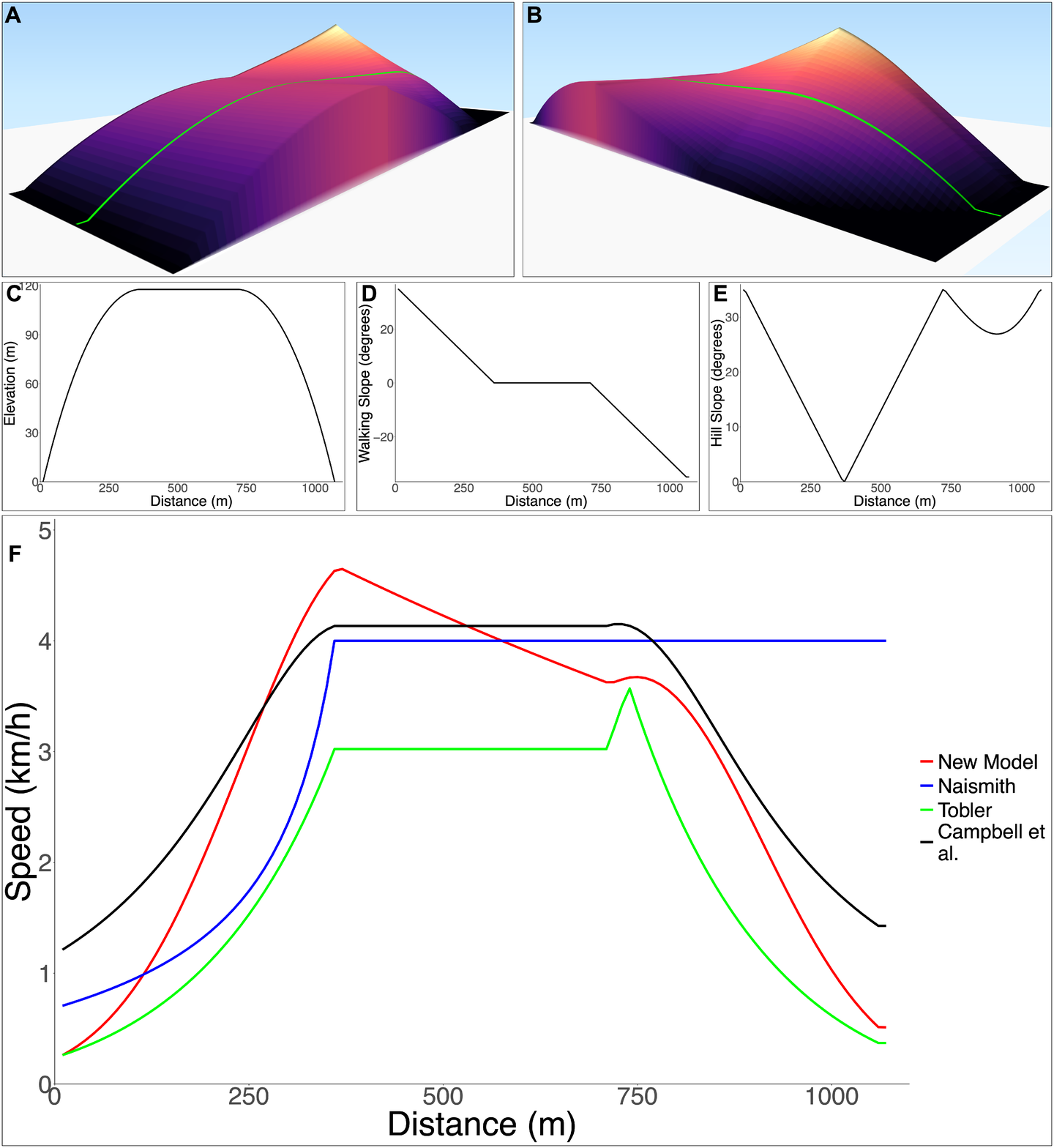}
    \captionsetup{width=1\linewidth}
    \caption[width=\textwidth]{{\bf Comparison of walking speed changes while crossing a simulated off-road terrain region.} (A), (B) The simulated route (green) across the terrain. Terrain is coloured by elevation value from low (dark) to high (light). (C) The elevation profile of the route, (D) The walking slope profile of the route, (E) The hill slope profile of the route. (F) Walking speed predictions for different models as the route is traversed. For Naismith's and Tobler's functions, the off-road variants of the models have been used. For the new model, the off-road unknown obstruction coefficients have been used.}
    \label{Fig17}   
\end{adjustwidth}
\end{figure}

\end{document}